\documentclass[12pt]{iopart}
\usepackage{iopams}
\usepackage{graphicx}
\usepackage[usenames,dvipsnames]{xcolor}
\usepackage{hyperref}
\usepackage{xspace}
\usepackage{enumitem}
\setlist[enumerate]{topsep=0pt,itemsep=0pt,partopsep=0pt,parsep=0pt}

\usepackage{cite}

\hypersetup{
    colorlinks=true,
    linkcolor=cyan,
    filecolor=magenta,      
    urlcolor=blue,
    citecolor=blue,
}

\usepackage{xcolor}
\definecolor{codegreen}{rgb}{0,0.6,0}
\definecolor{codegray}{rgb}{0.5,0.5,0.5}
\definecolor{codepurple}{rgb}{0.58,0,0.82}
\definecolor{backcolour}{rgb}{0.95,0.95,0.92}

\begin{document}
\title[Magnetohydrodynamics with Physics Informed Neural Operators]{Magnetohydrodynamics with Physics Informed Neural Operators}

\author{Shawn G. Rosofsky$^{1,2,3}$, E.~A. Huerta$^{1,2,4}$}
\address{$^1$ Data Science and Learning Division, Argonne National Laboratory, Lemont, Illinois 60439, USA}
\address{$^2$ Department of Physics, University of Illinois at Urbana-Champaign, Urbana, Illinois 61801, USA}
\address{$^3$ NCSA, University of Illinois at Urbana-Champaign, Urbana, Illinois 61801, USA}
\address{$^4$ Department of Computer Science, The University of Chicago, Chicago, Illinois 60637, USA}

\ead{elihu@anl.gov}

\vspace{10pt}
\begin{indented}
    \item[]\today
\end{indented}

\begin{abstract}
The modeling of multi-scale and 
multi-physics complex systems typically involves the 
use of scientific software that can optimally 
leverage extreme scale computing. Despite major developments 
in recent years, these simulations continue to be 
computationally intensive and time consuming. 
Here we explore the use of AI 
to accelerate the modeling of complex systems 
at a fraction of the computational cost of 
classical methods, and 
present the first application of 
physics informed neural operators to model 2D 
incompressible magnetohydrodynamics simulations. 
Our AI models incorporate tensor Fourier neural 
operators as their backbone, which we implemented 
with the \texttt{TensorLY} package. Our results 
indicate that physics informed neural operators can 
accurately capture the physics of magnetohydrodynamics 
simulations that describe laminar flows with 
Reynolds numbers $Re\leq250$. We also explore the 
applicability of our AI surrogates for turbulent 
flows, and discuss a variety of methodologies that may be 
incorporated in future work to create AI models that 
provide a computationally efficient and high fidelity 
description of magnetohydrodynamics simulations for 
a broad range of Reynolds numbers. The scientific software 
developed in this project is released with 
this manuscript. 
\end{abstract} 

\maketitle

\section{Introduction}
\label{ch6:sec:intro}
Turbulence emerges from laminar flow due to instabilities, has 
many degrees of freedom, and is commonly found in 
fluids with low viscosity. Its time-dependent and stochastic nature 
make it an ideal sandbox to explore whether AI methodologies 
are capable of learning and describing nonlinear phenomena 
that manifests from small to large scales. While the Navier-Stokes 
equations may be used to study flows of non-conductive fluids, 
flows of ionized plasmas present in astrophysical phenomena
may be considered perfectly conducting. These flows may be described 
by magnetohydrodynamics (MHD) equations, i.e., equations 
with currents, magnetic fields and the Lorentz force.

Theoretical and numerical modeling of MHD turbulence is 
critical to understand a variety of natural phenomena, encompassing 
astrophysical systems~\cite{Beresnyak2019:mhd_turbulence_review}, 
plasma physics~\cite{schekochihin2022:mhd_turbulence_review}, 
and geophysics~\cite{Pouquet2019:mhd_turbulence_review}. 
Some areas of interest in which MHD turbulence plays a 
crucial role are binary neutron star (BNS) 
mergers~\cite{Kuichi2015:BNS_KHI_highres}, black hole 
accretion, 
and supernova explosions~\cite{Beresnyak2019:mhd_turbulence_review}.  
The inherent complexity of MHD turbulence makes the modeling 
of these 
systems extremely difficult.

One of aspects of MHD turbulence responsible for such difficulty 
is the MHD dynamo, which amplifies the magnetic fields by converting 
kinetic energy into magnetic energy, starting at the smallest 
scales~\cite{Beresnyak2012:dynamo,Beresnyak2015,Beresnyak2019:mhd_turbulence_review,Grete2017}. In some cases, the MHD dynamo can produce 
amplification several orders of magnitude greater than that of 
the original fields~\cite{Kuichi2015:BNS_KHI_highres,Beresnyak2019:mhd_turbulence_review}. 
To resolve this magnetic field amplification, one must run simulations at the smallest of scales, where the MHD dynamo is most efficient \cite{Beresnyak2012:dynamo}. This scale is set by the Reynolds number, $Re$, of the flow. The higher the $Re$, the smaller the scale of the 
most efficient MHD dynamo amplification. However, astrophysical 
simulations in particular are at such high Reynolds numbers that 
it would be unfeasible to fully resolve such turbulence~\cite{Grete2017}.  Therefore, we must look to alternative ways to resolve such turbulent effects. One method is to approach MHD like a large eddy simulation 
(LES).  In LES, one ensures they possess sufficient resolution to 
resolve the largest eddies and employs a subgrid-scale (SGS) models 
to resolve turbulence at smaller scales. Some recent works~\cite{Grete2015,Grete2016,Vlaykov2016,Grete2017a,Grete2017b,Kessar2016,Aguilera-Miret2022:LES_BNS,Palenzuela2022:LES_BNS,Vigano2019:MHD_LES,Carrasco2020:MHD_LES_SR,Vigano2020:MHD_LES_grad_GR,Radice2020:BNS_SGS} 
have adopted traditional LES style SGS models for MHD simulations. 
Other works~\cite{Rosofsky2020a:SGS_MHD_2D,Karpov2022:supernova_ANN_LES} have examined the use of deep learning models as the LES model 
in MHD simulations. These deep learning models can use data to learn 
MHD turbulence properties not present in the traditional LES models, 
but are still in their early stages of development.

Another approach consists of accelerating scientific software used to 
model multi-scale and multi-physics simulations with AI surrogates. 
Neural operators are a very promising class of deep learning 
models that can accurately describe complex simulations 
at a fraction of the time and computational cost of traditional 
large scale simulations~\cite{kovachki2021neuraloperator,kovachki2021neuraloperator}.
Recent studies have employed neural operators to model turbulence 
in hydrodynamic simulations~\cite{Peng2022:FNO_3D_turbulence,Peng2022:FNO_attention,Li2022:FNO_LES}. In one study, the neural operators were compared to 
traditional LES style SGS models and were found to outperform the LES 
models in both accuracy and speed \cite{Li2022:FNO_LES}.

Physics informed neural operators (PINOs) incorporate physical 
and mathematical principles into the design, training and optimization 
of neural operators~\cite{PINO_li_anima}. It has been reported in 
the literature that this approach accelerates the convergence and 
training of AI models, and in some cases enables zero-shot learning~\cite{Rosofsky2022:PINO}. 
Several studies have illustrated the ability of PINOs to 
numerically solve partial differential equations (PDEs) that 
describe many complex problems~\cite{PINO_li_anima,Rosofsky2022:PINO}.

To further advance this line of research, 
in this paper we introduce  
AI surrogates to model multi-scale and 
multi-physics complex systems that are 
described by MHD equations. To this end, we produced 2D incompressible MHD 
simulations spanning a broad range of Reynolds numbers. 
We then trained PINOs with this data and evaluated them on 
a subset of simulations not observed in training.  Specifically, 
we compared the PINO predictions and the simulation values 
as well as their kinetic and magnetic energy spectra. 
\textit{To the best of our knowledge, this is the first study seeking to reproduce entire 
MHD simulations with AI models.}  As such, we focused 
our efforts on finding the strengths and weaknesses of 
this approach rather than optimizing our models as much 
as possible.  In doing so, 
this work provides a foundation of AI surrogates 
for MHD that future researchers will improve upon.

We organize this work as follows.  In Section~\ref{ch6:sec:mhd_sim} 
we introduce the incompressible MHD equations, and describe the 
numerical methods used to generate MHD simulation data. Then, we 
describe PINOs, and how we used them to solve MHD equations  
in Section~\ref{ch6:sec:PINO_modeling}. Section~\ref{ch6:sec:methods} details 
the methods we followed to create our PINOs, including 
generation of random initial data, model architecture, 
training procedure, evaluation criteria, and a description 
of our computational resources.  We present the results 
in Section~\ref{ch6:sec:results}. Final remarks and 
future directions of work are presented in 
Section~\ref{ch6:sec:conclusions}.

\section{Simulating Incompressible MHD}
\label{ch6:sec:mhd_sim}
\subsection{Equations}
\label{ch6:sec:mhd_equations}
The goal of this work is to reproduce the incompressible MHD equations with PINOs.  The incompressible MHD equations represent an incompressible fluid in the presence of a magnetic field $\mathbf{B}$. These equations are given by

\begin{eqnarray}
    \label{ch6:eq:vel}
    \partial_t \mathbf{u}+\mathbf{u} \cdot \nabla \mathbf{u} = -\nabla \left( p+\frac{B^2}{2} \right)/\rho_0 +\mathbf{B} \cdot \nabla \mathbf{B}+\nu \nabla^2 \mathbf{u}, \\
    \label{ch6:eq:mag}
    \partial_t \mathbf{B}+\mathbf{u} \cdot \nabla \mathbf{B} = \mathbf{B} \cdot \nabla \mathbf{u}+\eta \nabla^2 \mathbf{B}, \\
    \label{ch6:eq:div_vel}
    \nabla \cdot \mathbf{u} = 0, \\
    \label{ch6:eq:div_B}
    \nabla \cdot \mathbf{B} = 0,
\end{eqnarray}

\noindent where $\mathbf{u}$ is the velocity field, $p$ is 
the pressure, $B$ is the magnitude of the magnetic field, $\rho_0=1$ 
is the density of the fluid, $\nu$ is the kinetic viscosity, 
and $\eta$ is the magnetic resistivity.  We have two equations 
for evolution and two constraint equations.

To ensure the zero velocity divergence condition of 
Equation~\ref{ch6:eq:div_vel}, the pressure is typically computed 
in such a way that this condition always holds true. 
This is generally done by solving a Poisson equation 
to calculate the pressure at each time step.

For the magnetic field divergence of Equation~\ref{ch6:eq:div_B}, we 
lack any additional free parameters to ensure that the equation 
holds true at all times. There are several ways to ensure that 
this condition holds including hyperbolic divergence 
cleaning \cite{Dedner2002} and constrained 
transport \cite{Mocz2014:constrained_transport}.  For this work, 
we decide to preserve the magnetic field's zero divergence 
condition by instead evolving the magnetic vector potential 
$\mathbf{A}$.  This quantity is defined such that

\begin{eqnarray}
    \label{ch6:eq:A}
    \mathbf{B} = \nabla \times \mathbf{A},
\end{eqnarray}

\noindent which ensures that the divergence of $\mathbf{B}$ is zero 
to numerical precision as the divergence of a curl of a vector 
field is zero. By evolving magnetic vector potential $\mathbf{A}$ 
instead of the magnetic field $\mathbf{B}$, we have a new 
evolution equation for the vector potential $\mathbf{A}$. 
This equation is given by

\begin{eqnarray}
    \label{ch6:eq:vec_pot}
    \partial_t \mathbf{A} + \mathbf{u} \cdot \nabla \mathbf{A}=\eta \nabla^2 \mathbf{A}.
\end{eqnarray}

\subsection{Numerical Methods}
\label{ch6:sec:num_methods}

To simulate the MHD equations, we employed the Dedalus 
code~\cite{Burns2020:Dedalus}, an open source parallelized 
spectral python package that is designed to solve general PDEs. To 
obtain interesting results without additional computational difficulty, 
we elected to solve the incompressible MHD equations in 2D with 
periodic boundary conditions (BCs). This results in us solving 
a total of 3 evolution PDEs at each timestep--2 for the velocity 
evolution and 1 for the magnetic vector potential evolution. 
To visualize and more easily diagnose problems with the simulations, 
we include an additional PDE to evolve tracer particles denoted by $s$.  
The tracer particles had an evolution equation given by 

\begin{eqnarray}
    \label{ch6:eq:tracer}
    \partial_t s + \mathbf{u} \cdot \nabla s = \nu \nabla^2 s.
\end{eqnarray}

\noindent Moreover, we employed a pressure gauge of 
$\int{p \, dx^2} = 0$ for our equation. Numerically, most simulations 
were carried out on a unit length square grid at resolutions with 
a total number of grid points $N=128^2$, with $N_x=N_y=128$ 
points in the $x$ and $y$ directions, respectively. We also examined some 
lower resolution simulations with $N=64^2$ and 
higher resolution simulations with $N=256^2$
to explore how resolution may affect the results of the simulations.

As we have periodic BC, we employed a Fourier basis in each direction. 
To avoid aliasing error, we used a dealias factor of $3/2$ for 
these transformations. For timestepping, we used the RK4 integration 
method with a constant timestep of $\Delta t=0.001$. Simulation 
outputs were stored every 10 timesteps resulting in us recording 
data at interval $t=0.01$ time units.  We ran these simulations 
until a time of $t=1$. For each set of parameters, we produced 1,000 simulations, each with different
initial data.

\subsection{Reynolds Number}
\label{ch6:sec:Re}

An important consideration in this study was how the Reynolds number 
affects simulations and our AI model's ability to reproduce the 
results of the simulations. The Reynolds number is a dimensionless 
quantity that expresses the ratio of inertial to viscous forces. 
The higher the Reynolds number is, the more prevalent the effects 
of small scale phenomena are to the simulation. This often gives 
rise to turbulence at high Reynolds numbers. For systems described by 
MHD equations, there are actually 2 types of Reynolds numbers of 
interest.  They are the kinetic Reynolds number, $Re$, and 
the magnetic Reynolds number, $Re_m$. We define the 
quantities in our simulation such that these  Reynolds 
numbers are given by 

\begin{eqnarray}
    \label{ch6:eq:Re}
    Re &= 1/\nu, \\
    \label{ch6:eq:Re_m}
    Re_m &= 1/\eta.
\end{eqnarray}

\noindent The ratio between these two Reynolds numbers is called the magnetic Prandtl number $Pr_m$ defined as

\begin{eqnarray}
    \label{ch6:eq:Pr_m}
    Pr_m = \frac{Re_m}{Re} = \frac{\nu}{\eta}.
\end{eqnarray}

In this study, we sought to characterize how the Reynolds number 
affects our models and determine if there is a cutoff Reynolds number, 
after which point, the models' performance degrades considerably. 
We generally kept $Re=Re_m$ or, equivalently, $Pr_m=1$ unless 
otherwise noted and looked at Reynolds number values of 100, 250, 
500, 750, 1,000, and 10,000.  

One difference between MHD and hydrodynamic simulations is that, 
compared to the pure hydrodynamic case, much of the magnetic 
field energy is stored at high wavenumbers which occur at smaller 
scales. Thus, the models must be able to characterize high 
frequency features if they are to successfully reproduce the 
results of the magnetic field evolution.

\section{Modeling MHD with PINOs}
\label{ch6:sec:PINO_modeling}

\subsection{PINOs}
\label{ch6:sec:PINO}
The goal of a neural operator (NO) is to reproduce the results 
of an operator given some input fields using neural networks. 
A common class of operator often studied by NOs are PDEs.  
To model PDEs, NO take the coordinates, initial conditions 
(ICs), BCs, and coefficient fields as inputs. As outputs, NOs provide 
the solution of the operator, in this case the PDE, at the 
desired spacetime coordinates.

There are a variety of different NOs that have been studied in 
recent works.  These include DeepONets, physics informed DeepONets, 
low-rank neural operators (LNO), graph neural operators (GNO), 
multipole graph neural operators (MGNO), Fourier neural operators 
(FNO), factorized Fourier neural operators (fFNO), 
and PINO~\cite{lu2021learning,wang2021learning, Li2020GraphNO,Li2020MGNO,Li2021FourierNO,kovachki2021neuraloperator,Tran2021:fFNO,PINO_li_anima,Rosofsky2022:PINO}. In this study we use PINOs.

PINOs are a generic class of NOs that involve using an existing 
neural operator and employing physics informed methods in the 
training~\cite{PINO_li_anima,Rosofsky2022:PINO}.  Physics informed 
deep learning methods involves encoding known information about 
the physical system into the model during training~\cite{raissi2017physicsI,raissi2017physicsII,raissi2019physics,Pang2019fPINNs,lu2021deepxde,Karniadakis2021:physics_informed_machine_learning}.  This physics information may include governing PDEs, 
symmetries, constraint laws, ICs, and BCs.  By including 
such physical knowledge, physics informed methods enable 
better generalization of the results of deep learning 
models~\cite{Karniadakis2021:physics_informed_machine_learning}. 
In systems where we have a lot of knowledge about the physical system 
like PDEs, such methods are especially useful as theoretically 
we could learn from just this physics knowledge.  In practice, we 
add data to help PINOs converge to the correct solution more quickly.

One of the most common ways of encoding this physics into a 
neural network model is to incorporate physics information 
into the loss function~\cite{raissi2017physicsI,raissi2017physicsII,raissi2019physics,Pang2019fPINNs}.  In other words, violations of these physics laws 
appear as terms in the loss function that are reduced over time 
during training. Thus, this technique treats physics laws as 
soft constraints learned by the neural network as it trains.

For the backbone NO, we selected a variant of 
fFNO~\cite{Tran2021:fFNO} that employed the 
\texttt{TensorLY}~\cite{tensorly} package to perform tensor 
factorization.  We call this model a tensor Fourier neural operator 
(tFNO).  The base FNO model~\cite{Li2021FourierNO} applies the 
fast Fourier transform (FFT) to the data to separate it into 
its component frequencies and apply its weights before performing 
an inverse FFT to convert back to real space. One particular feature 
of the FNO is its ability to perform zero-shot super-resolution, 
in which the model predicts on higher resolution data than it was 
trained on~\cite{Li2021FourierNO,kovachki2021neuraloperator}.  We 
hope such properties of the FNO model would allow it to learn 
the high frequency properties of the MHD equations.  

To further augment the model, we utilized factorization 
with \texttt{TensorLY} within the spectral layers of the model, 
so that it becomes an tFNO.  The factorization was found to 
significantly improve the generalization of the neural network 
in the initial tests.  The decision to use \texttt{TensorLY} was 
inspired by experimental code found in the \texttt{GitHub} 
repository of~\cite{PINO_li_anima}.  We will describe the model 
architecture in more detail in Section~\ref{ch6:sec:architecture}.

\subsection{Applying to MHD Equations}
\label{ch6:sec:physics_info}

Now let us discuss in more detail how to apply physics informed 
methods to model the MHD equations with neural operators. There 
are several aspects to modeling the MHD equations with such methods 
as denoted by each loss term.  These are the data loss 
$\mathcal{L}_{\textrm{data}}$, the PDE loss $\mathcal{L}_{\textrm{PDE}}$, 
the constraint loss $\mathcal{L}_{\textrm{constr}}$, the IC 
loss $\mathcal{L}_{\textrm{IC}}$, and the BC loss $\mathcal{L}_{\textrm{BC}}$.

The data loss is modeled by obtaining simulation data and ensuring 
that the PINO output matches the simulation results.  This requires 
us to produce a large quantity of simulations to provide 
sufficient training data like those described in 
Section \ref{ch6:sec:num_methods} to span the solution space.  
We will discuss in more detail how we produced a sufficient 
number of simulations in Section~\ref{ch6:sec:data_gen}. 
Then we use the relative mean squared error (MSE) between 
the simulation and the PINO predictions to define the value 
of the data loss $\mathcal{L}_{\textrm{data}}$.

The PDE loss describes the violations of the time evolution 
PDEs of the PINO outputs. Specifically, we encode these 
time evolution PDEs as part of the loss function. This 
requires us to represent the spatial and temporal derivatives of 
the output fields. Although some NOs are able to employ the 
automatic differentiation of the deep learning framework to 
calculate such derivatives~\cite{wang2021learning}, this method 
is too memory intensive for the tFNO architecture. Instead, we 
employ Fourier differentiation to represent the spacial derivatives, 
which computes highly accurate derivative values while 
conveniently assuming periodic BC that exists in our problem. 
For time derivatives, we use second order finite differencing. 
We note that such a time differencing technique cannot be used 
during the original simulation as this requires knowledge of 
future times to compute the time derivative of the current time.

Specifically, the PDEs we modeled with this technique were 
Equation~\ref{ch6:eq:vel} for velocity evolution and 
Equation~\ref{ch6:eq:A} for magnetic potential evolution. 
In this case, the model has a total of 3 output fields in 2D, 
the velocity in the $x$ direction $u$, the velocity in the $y$ 
direction $v$, and the magnetic potential $A$. We also tested 
replacing the magnetic vector potential evolution with the 
magnetic field evolution of Equation \ref{ch6:eq:mag}, 
which results in 4 output fields in 2D.  The fields in this 
case are the velocity in the $x$ direction $u$, the velocity 
in the $y$ direction $v$, the magnetic field in the $x$ 
direction $B_x$, and the magnetic field in the $y$ direction $B_y$. 
We note that in testing, the former representation of the 
MHD equations produced better results.  The PDE loss 
$\mathcal{L}_{\textrm{PDE}}$ is then defined as the MSE loss between 
zero and the PDE, after putting all the terms on the same 
side of the equation.

Due to the complexity of both representations of the MHD equations, 
we tested the PDE loss on the data produced by the simulations. 
We found that the loss was small, albeit nonzero. This nonzero 
PDE loss was expected since the numerical methods for computing 
the derivatives during the simulation differs from those of 
computing the derivatives of the PDE during the loss function. 
Some particular notable differences are having fewer time steps 
in the output data than was used during the simulation, 
using different time derivative methods (e.g., RK4 during simulation 
vs second order finite differencing during the loss computation), 
and lacking dealiasing in the spatial derivatives in the PDE 
loss function.

The constraint loss illustrates the deviations of the 
elliptic constraint equations of the MHD equations. 
Specifically, these refer to the velocity divergence free 
condition of Equation~\ref{ch6:eq:div_vel} and the magnetic 
divergence free condition of Equation~\ref{ch6:eq:div_B}. 
We implemented these constraints similarly to the time 
evolution equations in the PDE loss, but without any 
time derivative terms. The constraint loss $\mathcal{L}_{\textrm{constr}}$ 
is then the MSE between each of the constraint equations and zero.  
We note that we are representing the magnetic fields by the 
magnetic potential, the magnetic field divergence free 
condition is satisfied up to the numerical precision of 
the curl operator regardless of the prediction of the neural 
network. However, we still include the term for completeness 
and this condition is not guaranteed to be nonzero if we are 
trying to compute the magnetic fields directly.

The IC loss tells the model to associate the input field with 
the output at $t=0$.  Although this function can often be 
achieved with the data loss alone, the IC loss emphasizes 
the importance of predicting the correct IC and enables training 
in the absence of data.  Both the significance of the IC and 
the ability to train without data stem from the PDE loss 
term. Theoretically, one can train by correctly predicting 
the IC, the BCs, and evolving the PDE correctly forward 
in time, although in practice, data helps the model converge 
more quickly.  However, an incorrect IC results in the PDE 
evolving the wrong data forward in time. Thus, we add in the IC 
as its own term to encourage the model to first compute the IC correctly before learning the time evolution later in training. 
We calculate the IC loss $\mathcal{L}_{\textrm{IC}}$ by taking the 
input fields and computing the relative MSE between said 
input fields and the outputs fields at $t=0$.

Finally, the BC loss $\mathcal{L}_{\textrm{BC}}$ describes the violations 
of the boundary terms. In our case, the tFNO model architecture 
ensures that we have the desired periodic BC. Therefore, we do 
not use such a term in our model. However, we mention this term 
for generality because not all BCs and the periodicity of 
the tFNO architecture can be removed by zero padding the 
inputs along the desired non-periodic axis.

In physics informed deep learning methods, one must be careful 
when combining fields of different magnitudes to ensure they 
all have an equal contribution to the loss. Therefore, our model 
has the ability to normalize the input fields and denormalize the 
output fields by multiplying by predetermined constants. 
Moreover, we assign a weight to each term when combining 
loss terms for different fields and equation. We add additional 
weights when combining different losses.  Thus, our 
loss $\mathcal{L}$ is given as 

\begin{eqnarray}
    \label{ch6:eq:loss}
    \mathcal{L} &= w_{\textrm{data}} \mathcal{L}_{\textrm{data}} + w_{\textrm{PDE}} \mathcal{L}_{\textrm{PDE}} + w_{\textrm{constr}} \mathcal{L}_{\textrm{constr}} 
    + w_{\textrm{IC}} \mathcal{L}_{\textrm{IC}}\,,
\end{eqnarray}

\noindent where $w_{\textrm{data}}$ is the data weight, 
$w_{\textrm{PDE}}$ is 
the PDE weight, $w_{\textrm{constr}}$ is the constraint weight, 
and $w_{\textrm{IC}}$ is the IC weight.

\begin{figure*}
    \centering
    \includegraphics[width=\textwidth]{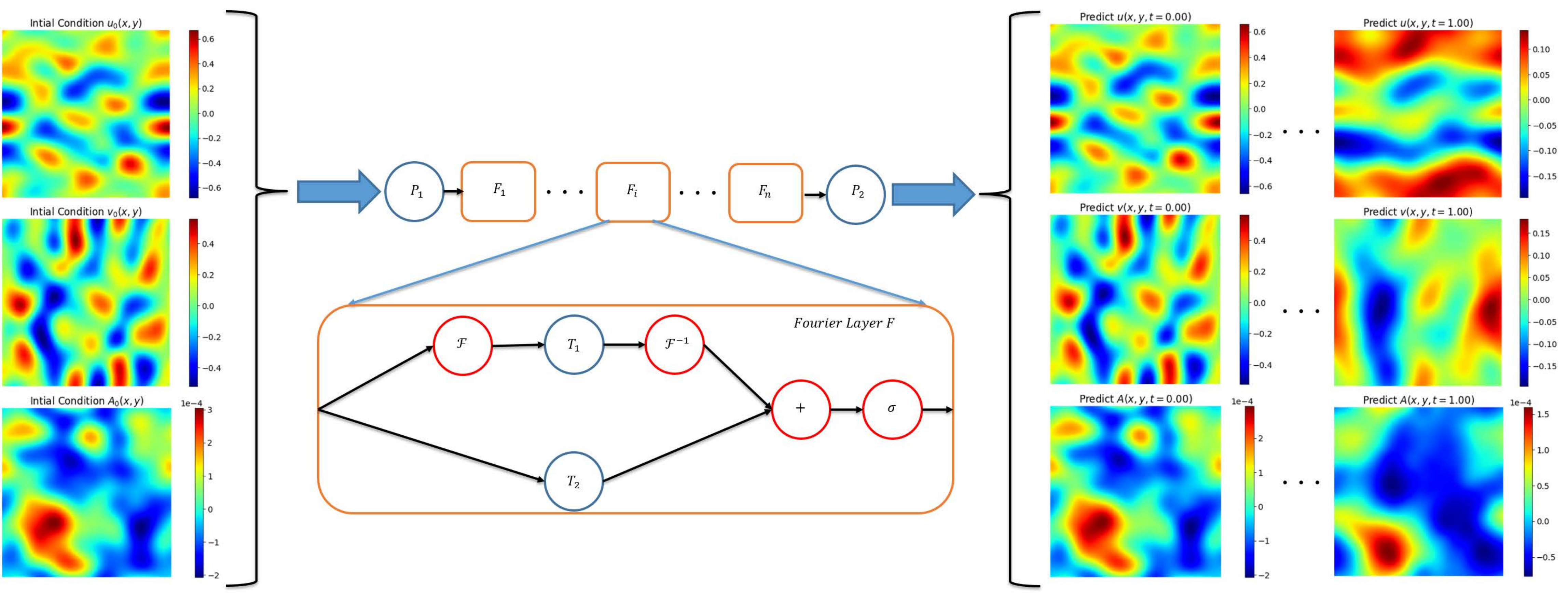}
    \caption{\textbf{Architecture:} Schematic representation of 
    the architecture of our PINO models, which use the 
    \texttt{TensorLY}~\cite{tensorly} package to perform tensor 
    factorization, thereby using tensor Fourier neural 
    operators (tFNO) as the backbone of our AI models. On 
    the far left, we display sample input fields that are fed 
    into our PINO, which are composed of the $u$, $v$, and $A$ 
    initial conditions (ICs). To the right of the input 
    fields, we illustrate 
    how the data is transformed as it goes through the model. 
    First, this input data is lifted into a higher 
    dimension representation by the neural network, 
    $P_1$. Then, the data 
    enters a series of Fourier layers labeled 
    $\{F_1, ..., F_i, ..., F_n\}$.  The inset zooms into one 
    of these Fourier layers.  In the inset, labelled as 
    "\textit{Fourier Layer F}", we apply a series 
    of operations that consist of non-local integral operators 
    and nonlinear activation functions. Specifically, $T_1$ 
    represents a linear transform that employs CP decomposed 
    tensors as weights, $T_2$ represents a local linear 
    transform, and $\sigma$ represents a nonlinear 
    activation function. $\mathcal{F}$ and $\mathcal{F}^{-1}$ 
    stand for Fourier transform and inverse Fourier 
    transform, respectively. Eventually, the neural 
    network $P_2$ projects back into position space, 
    producing the output shown on the right panels, 
    which describe the time evolution of the system. 
    These outputs are presented on the right of the 
    image for the output fields $u$, $v$, and $A$ 
    at times that range from $t=0$ to $t=1$.  }
    \label{ch6:fig:architecture}
\end{figure*}

\begin{figure*}[htbp]
    \centering
    \includegraphics[width=\textwidth]{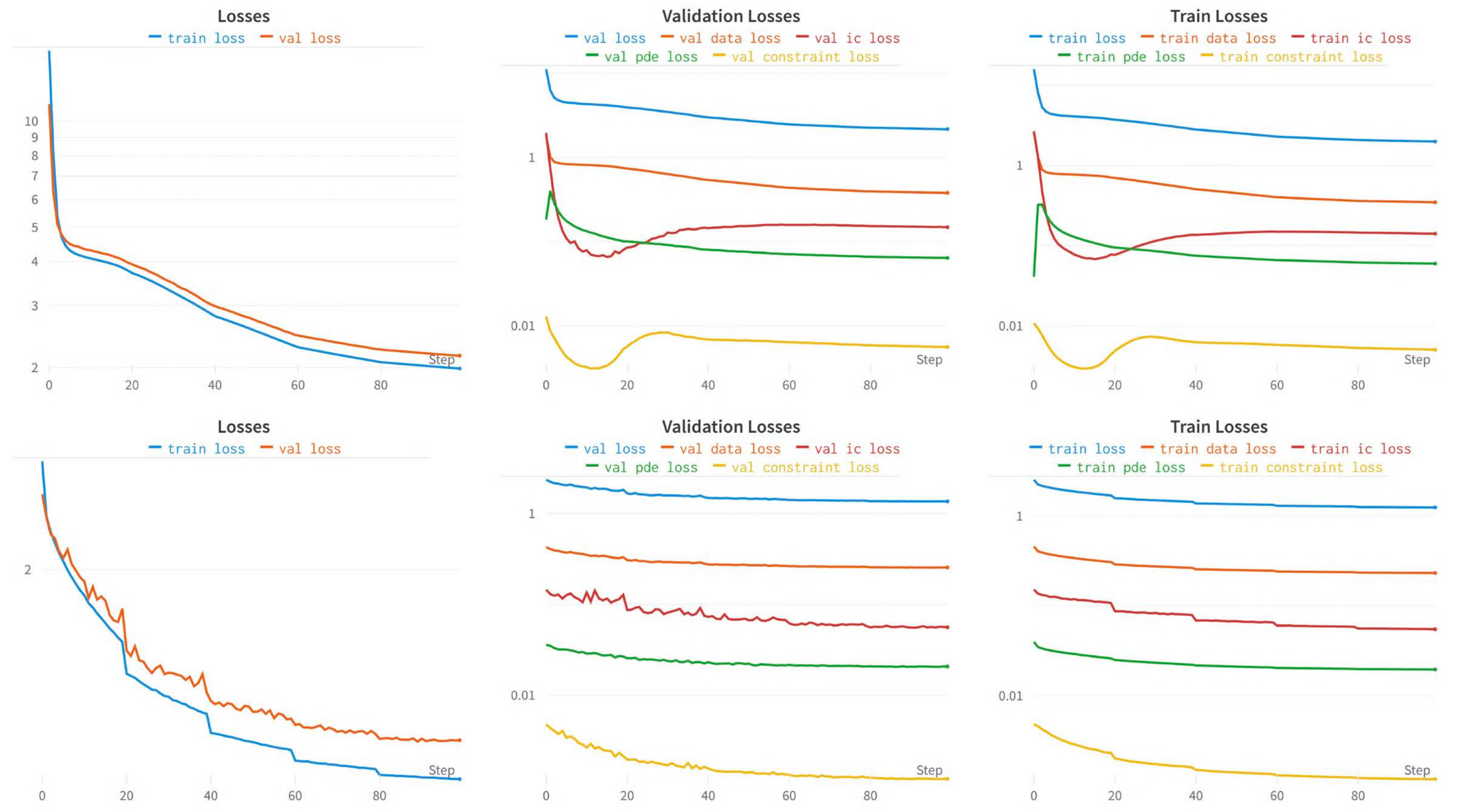}
    \caption{\textbf{Loss Curves:} Here we display the loss curves 
    for the training.  The top row depicts the original $Re=100$
    case that was trained from scratch.  The final checkpoint that
    was created by this run was used as the starting point for all
    models presented in this study.  The bottom row shows an
    $Re=250$ model that was trained starting at said checkpoint
    The left panels depicts the total loss 
    for the training and validation data on a linear scale. 
    The center panels shows the training losses for each 
    of the different types of losses we track--total loss, 
    data loss, PDE loss, IC loss, and constraint loss. The 
    right panels display these same losses, but for 
    the validation data. To illustrate the different scales 
    of these losses, the center and right panels use a log scale 
    for the y-axis.  The decreases in loss every 20 epochs 
    occur due to the scheduler decreasing the learning rate 
    by a factor of 2 at those epochs.  
    }
    \label{ch6:fig:losses}
\end{figure*}

\section{Methods}
\label{ch6:sec:methods}
\subsection{Data Generation}
\label{ch6:sec:data_gen}

To generate the training data, we first needed to produce 
initial data before running it in the simulations described 
in Section~\ref{ch6:sec:num_methods}.  These initial data fields are 
produced using Gaussian random field (GRF) method similar 
to~\cite{Rosofsky2022:PINO}, in which the kernel was transformed 
into Fourier space to obey our desired periodic BCs.  Specifically, 
we used the radial basis function kernel (RBF) to produce smooth 
initial fields. This kernel $k_l$ is defined as

\begin{eqnarray}
    \label{ch6:eq:grf}
    k_{l}\left(x_{1}, x_{2}\right)=\exp\left(-\frac{\left\|x_{1}-x_{2}\right\|^{2}}{2 l^{2}}\right),
\end{eqnarray}

\noindent where $l$ is the length scale of typical spatial 
deviations in the data.  We used $l=0.1$ for all fields in 
this works unless otherwise stated. We also needed to ensure that 
the velocity and magnetic fields are both divergence free. 
Therefore, we produced 2 initial data fields, the vorticity 
potential, $\psi$, and the magnetic potential $A$.  We defined 
$\psi$ such that

\begin{eqnarray}
    \label{ch6:eq:psi}
    \mathbf{v} = \nabla \times \psi,
\end{eqnarray}

\noindent which guarantees the velocity fields are divergence 
free initially. The magnetic potential $A$ is defined 
in Equation~\ref{ch6:eq:vec_pot} in a similar manner to prevent the 
presence of divergences in the initial magnetic fields.

We multiplied the resulting initial data fields by a constant to 
ensure that the resulting fields have appropriate magnitude and 
are numerically stable. For example, we need to prevent 
Courant-Friedrichs-Lewy (CFL) condition violations, which can 
occur if the velocities are too high for a given resolution and 
timestep choice~\cite{Courant1928:CFL}.  Numerical instabilities 
can also occur if the initial magnetic potential values are too 
high and may cause the simulation to fail. We chose these constants 
to be $c_{\psi}=0.1$ for the vorticity potential 
$\psi$ and $c_{A}=0.005$ for the magnetic potential $A$.

\subsection{Model Architecture}
\label{ch6:sec:architecture}

A schematic of the tFNO models utilized in this study is 
presented in Figure~\ref{ch6:fig:architecture}. The size and dimensions 
of the model can be described by 3 hyperparameters--the width, 
the number of Fourier modes, and the number of layers. We used 
4 layers for all models in this work. The width and number of 
Fourier modes were the same across all layers of the model 
in this work and were both set to 8 unless stated otherwise. 
Although we suspect that we could get better results by 
increasing the these hyperparameters, especially the number 
of Fourier modes which may have helped the models reproduce 
small scale features in the flow, doing so proved too memory 
intensive for our GPUs.

We factorized the weight tensors within the spectral layers 
with \texttt{TensorLY} to improve generalizability. 
Specifically, we used a canonical polyadic tensor 
decomposition \cite{erichson2020:randomized_CP} with a 
rank of 0.5 to perform this factorization. Prior to producing 
the outputs, 
the models employ a fully connected layer of width 128.

In addition, the model normalizes its data internally by 
dividing by a constant input normalization factor to ensure 
that magnitude varying inputs are treated the same way. 
In particular, the magnitude of the magnetic potential $A$ 
was considerably less than that of the velocity fields $u$ 
and $v$. Similarly, we multiplied the output by a constant 
output normalization factor to alleviate the tFNO model 
from having to produce results with significantly different magnitudes. 
For this work, we used the same value for the 
input and output normalization fields of $1$ for the velocity 
fields and $0.00025$ for the magnetic potential $A$. Finally, 
we selected the Gaussian error linear unit (GELU) 
for the nonlinearity of these models \cite{hendrycks2016GELU}.

\subsection{Training}
\label{ch6:sec:training}
To accelerate the training, we would employ transfer learning 
across different Reynolds number $Re$. Specifically, we began by training a model at a resolution
$N=128^2$ at $Re=100$ from scratch.  The checkpoint generated
from this run was used as the starting point for all the
other models.   In turn, this transfer learning saved 
considerable time when training new models.  
We trained models initialized from the aforementioned checkpoint
at resolution of $N=128^2$ for $Re$ values of 
\(Re=\{100, 250, 500, 750, 1000, 10000\}\). 
We then compared these models to additional ones trained
at resolutions of $N=64^2$ and $N=256^2$.  These models were
for the same $Re$ as the previous $N=128^2$ resolutions, except
for $Re=10,000$.  All models trained using $950$ simulations 
that encompassed the training data
and were evaluated using the remaining $50$ simulations
that served as the test data.  In Figure~\ref{ch6:fig:losses}, 
we display some loss curves for the $Re=100$ model that
was trained from scratch and for the $Re=250$ model
both at resolutions of $N=128^2$.

For most models, we trained on all available timesteps.
However, we wanted to see the effect of being asked to
output fewer time steps had on the model.  Therefore,
we trained additional models that skipped several timesteps.
These trained at $Re=250$ and $N=128^2$ and are described
further in Section~\ref{ch6:sec:results}.

We trained these models using the \texttt{PyTorch} deep 
learning framework \cite{paszke2017:PyTorch}.
All models in this study trained for 100 epochs starting
from the pretrained checkpoint.  Most models used
an initial learning rate of $5\times 10^{-4}$.  However,
the models at $N=128^2$ with $Re$ values of $750$,
$1,000$, and $10,000$ had initial learning weights of
0.001.  The different initial learning rate for these runs 
was unintentional.  We suspect this higher may have
improved the performance of these runs based on 
how they perform compared to the same $Re$ runs at 
other resolutions, but no rigorous
study of learning rate optimization was performed.
We employed a scheduler to decrease the learning rate
by a factor of 2 every 20 epoch to help finetune the
models.  We selected an
AdamW optimizer \cite{loshchilov2017:AdamW} to optimize
the models.  This optimizer had a weight 
decay value of 0.1, $\beta_1=0.9$, $\beta_2=0.999$.

We set most of the weight hyperparameters of the various loss 
terms to $1$ with a few notable exceptions. We set the weight 
of the magnetic potential evolution equation loss $w_{\textrm{DA}}$ 
to $10^6$ as the magnitude of the term was small compared 
to that of the other evolution equations. In addition, we used 
a value of the constraint loss weight $w_{\textrm{constr}}$ of $10$. 
Finally, we selected $w_{\textrm{data}}=5$ as our data 
loss weight. We logged specific hyperparameters used in training each PINO 
using \texttt{WandB} \cite{wandb} for reproducibility. In addition, we ran an experiments where we
modified the PDE loss weight $w_{\textrm{PDE}}$ hyperparameter to explore its
impact on our AI models.  These were done at $Re=250$ and $Re=500$ both at $N=128^2$ and are described further in 
Section~\ref{ch6:sec:results}.

\subsection{Computational Resources}
\label{ch6:comp_resources}

We initially computed a subset of the results on the 
Pittsburgh Supercomputing Center's Bridges-2 cluster.
Specifically, we used the V100 GPUs on this cluster. 
We primarily used the 16 GB variant of these GPUs, though some cases utilized 
the 32 GB variation. To generate the training data, for the initial portion of the study, 
we utilized the CPUs on the Bridges-2 cluster's GPU nodes. We 
used a single GPU node for this process, which has two Intel 
Xeon Gold 6248 ``Cascade Lake'' CPUs, which have 20 cores, 
2.50-3.90GHz, 27.5 MB LLC, 6 memory channels each.
After verifying that our models worked as expected, 
we scaled up our experiments using NCSA's Delta cluster.
For training, we employed the cluster's NVIDIA A100 GPUs,
training each model on a single GPU.
To generate an expanded quantity of training data,
we employed the Delta cluster's CPU nodes.  
These come equipped with 128 AMD  EPYC 7763 ``Milan'' 
(PCIe Gen4) CPUs. We parallelized the generation of the 
training data such that multiple
simulations were generated at one, but each only using one core. 
We considered low (\(N=64^2)\), standard (\(N=128^2)\), and high resolution (\(N=256^2)\) simulations. 

\subsection{Evaluation Criteria}
\label{ch6:sec:eval_criteria}

To evaluate the performance of our PINO models, we look at 
the relative MSE.  This allows us to compare errors of the 
various fields despite them differing in magnitude.  We report 
the relative MSE for the predictions the PINOs on the test 
dataset for each field of interest. In addition, we compute 
the total relative MSE, $\textrm{MSE}_{tot}$, for our PINOs on this 
data which we defined as

\begin{eqnarray}
    \label{ch6:eq:mse_tot}
    \textrm{MSE}_{tot} = \textrm{MSE}_u + 
    \textrm{MSE}_v + \textrm{MSE}_A,
\end{eqnarray}

\noindent where $\textrm{MSE}_u$, $\textrm{MSE}_v$, 
and $\textrm{MSE}_A$ are the \textrm{MSE} 
values of the $u$, $v$, and $A$ fields respectively. 
In addition, we computed the kinetic energy spectra 
$E_{kin}(k)$ and magnetic energy spectra, $E_{mag}(k)$, of the 
PINO predictions and the ground truth simulations. This allowed 
us to compare how the models perform at various scales as 
specified by the wavenumber $k$.

\begin{table}
\caption{\textbf{Summary of Results:} The $Re$ column displays the Reynolds number of the model. The Resolution column provides the resolution of the model.  The $N_{train}$ and $N_{test}$ columns tell us how many simulations were used in training and testing respectively for each PINO.  The MSE $u$, MSE $v$, and MSE $A$ give us the relative MSE values of the $u$, $v$, and $A$ fields respectively.  The far right column, MSE Total, lists the sum of the MSE values of the $u$, $v$, and $A$ fields.}
\begin{indented}
\lineup
\item[]\begin{tabular}{@{}*{7}{l}}
\br                              
\textbf{\textbf{Standard Runs}}      & $\mathbf{Re}$                   & \textbf{Resolution}          & \textbf{MSE u}          & \textbf{MSE v}          & \textbf{MSE A}          & \textbf{MSE Total}          \\ 
\mr
                                     & 100                             & $128^2$                      & 0.019433                & 0.023787                & 0.039242                & 0.082462103                 \\
                                     & 250                             & $128^2$                      & 0.065276                & 0.070364                & 0.119925                & 0.255565267                 \\
                                     & 500                             & $128^2$                      & 0.136089                & 0.146443                & 0.262841                & 0.545372143                 \\
                                     & 750                             & $128^2$                      & 0.160067                & 0.168168                & 0.313788                & 0.642023818                 \\
                                     & 1000                            & $128^2$                      & 0.185853                & 0.19528                 & 0.362628                & 0.7437606                   \\
                                     & 10000                           & $128^2$                      & 0.280736                & 0.28256                 & 0.557567                & 1.120862162                 \\
                                     & 100                             & $64^2$                       & 0.018748                & 0.022466                & 0.03603                 & 0.077244182                 \\
                                     & 250                             & $64^2$                       & 0.062604                & 0.070628                & 0.122528                & 0.255759554                 \\
                                     & 500                             & $64^2$                       & 0.129726                & 0.139503                & 0.249703                & 0.518932519                 \\
                                     & 750                             & $64^2$                       & 0.170337                & 0.180869                & 0.325797                & 0.67700303                  \\
                                     & 1000                            & $64^2$                       & 0.196846                & 0.207568                & 0.374421                & 0.778835213                 \\
                                     & 100                             & $256^2$                      & 0.019498                & 0.023901                & 0.037558                & 0.080957422                 \\
                                     & 250                             & $256^2$                      & 0.064941                & 0.076371                & 0.130074                & 0.271386759                 \\
                                     & 500                             & $256^2$                      & 0.134618                & 0.147628                & 0.261671                & 0.543917469                 \\
                                     & 750                             & $256^2$                      & 0.176292                & 0.190424                & 0.339657                & 0.706372823                 \\
                                     & 1000                            & $256^2$                      & 0.203172                & 0.218052                & 0.389165                & 0.81038871                  \\
                                     \hline\hline
\textbf{$\mathbf{w_{\textrm{PDE}}}$} & \textbf{\textbf{$\mathbf{Re}$}} & \textbf{Resolution}          & \textbf{MSE u}          & \textbf{MSE v}          & \textbf{MSE A}          & \textbf{MSE Total}          \\
0                                    & 250                             & $128^2$                      & 0.065495                & 0.070718                & 0.120027                & 0.256239378                 \\
1                                    & 250                             & $128^2$                      & 0.065276                & 0.070364                & 0.119925                & 0.255565267                 \\
2                                    & 250                             & $128^2$                      & 0.06514                 & 0.070161                & 0.119934                & 0.25523497                  \\
5                                    & 250                             & $128^2$                      & 0.064882                & 0.069786                & 0.120038                & 0.254706307                 \\
10                                   & 250                             & $128^2$                      & 0.064688                & 0.069522                & 0.120324                & 0.254534433                 \\
0                                    & 500                             & $128^2$                      & 0.136623                & 0.147041                & 0.264064                & 0.547728002                 \\
1                                    & 500                             & $128^2$                      & 0.136089                & 0.146443                & 0.262841                & 0.545372143                 \\
2                                    & 500                             & $128^2$                      & 0.135796                & 0.146161                & 0.262429                & 0.544385653                 \\
5                                    & 500                             & $128^2$                      & 0.135441                & 0.14579                 & 0.262019                & 0.543249941                 \\
10                                   & 500                             & $128^2$                      & 0.135667                & 0.145995                & 0.262715                & 0.544377502                 \\
\hline\hline
\textbf{$\mathbf{t_{step}}$}         & \textbf{\textbf{$\mathbf{Re}$}} & \textbf{\textbf{Resolution}} & \textbf{\textbf{MSE u}} & \textbf{\textbf{MSE v}} & \textbf{\textbf{MSE A}} & \textbf{\textbf{MSE Total}} \\
1                                    & 250                             & $128^2$                      & 0.065276                & 0.070364                & 0.119925                & 0.255565267                 \\
2                                    & 250                             & $128^2$                      & 0.064159                & 0.069215                & 0.115976                & 0.249350436                 \\
3                                    & 250                             & $128^2$                      & 0.062734                & 0.067394                & 0.111937                & 0.242065011                 \\
4                                    & 250                             & $128^2$                      & 0.061139                & 0.065587                & 0.108474                & 0.235199714  \\
\br
\end{tabular}
\end{indented}
\label{ch6:tab:pino_results}
\end{table}

\section{Results}
\label{ch6:sec:results}

\noindent Here we present results for the accuracy with which our 
PINO models solve the MHD equations, and then compare these 
AI predictions with high performance computing MHD simulations. 
These simulations are evolved in the time domain \(t\in[0,1]\). 
We present results for a broad range of Reynolds number $Re$, 
summarized in Table~\ref{ch6:tab:pino_results}. Our PINO 
simulations and traditional MHD simulations are compared 
throughout their entire evolution, and then we take a 
snapshot of their evolution at \(t=1\). We do this to 
capture the largest discrepancy between AI-driven simulations 
and traditional PDE solvers at a time where numerical 
errors and other discrepancies between these methodologies 
are maximized. We examined a variety of factors that impact the performance of 
our models, including:
\begin{itemize}
    \item{\textbf{Reynolds number:}
    We quantify the ability of our AI surrogates to describe 
    the physics of MHD simulations for laminar and turbulent flows. 
    In general, simulations with higher Reynolds numbers, $Re$, 
    i.e., more turbulent flows, are more challenging to 
    describe accurately.}
    \item \textbf{Resolution:}
    We compare the performance of our models using three 
    grid resolutions: $64^2$, $128^2$, and $256^2$,
    which we denote as the low, standard, and high resolutions, 
    respectively.  In what follows, we present results for the standard 
    resolution, \(N=128^2\). Results for low and high 
    resolutions are presented in~\ref{sec:appex}.
    \item \textbf{PDE loss weight:}
    We looked at how changing the PDE loss weight $w_{\textrm{PDE}}$ impacts our AI models.
    This represents how much the physics informed aspect of the model improved the results.
    We tested $w_{\textrm{PDE}}$ values of 0, 1, 2, 5, and 10.
    \item \textbf{Timestep:}
    By default, our AI models use 101 timesteps to cover the 
    time range \(t\in[0,1]\). We use the quantity $t_{\textrm{step}}$
    to represent the frequency of sampling from the full set of times.
    For example, $t_{\textrm{step}}=1$ uses all timesteps and $t_{\textrm{step}}=2$ uses every other timestep.
    For this experiment, we used $t_{\textrm{step}}$ values of 1, 2, 5, and 10.
\end{itemize}

\subsection{Resolution}
\label{ch6:sec:Re_results}
\begin{figure*}[htbp!]
    \centering
    \includegraphics[width=\textwidth]{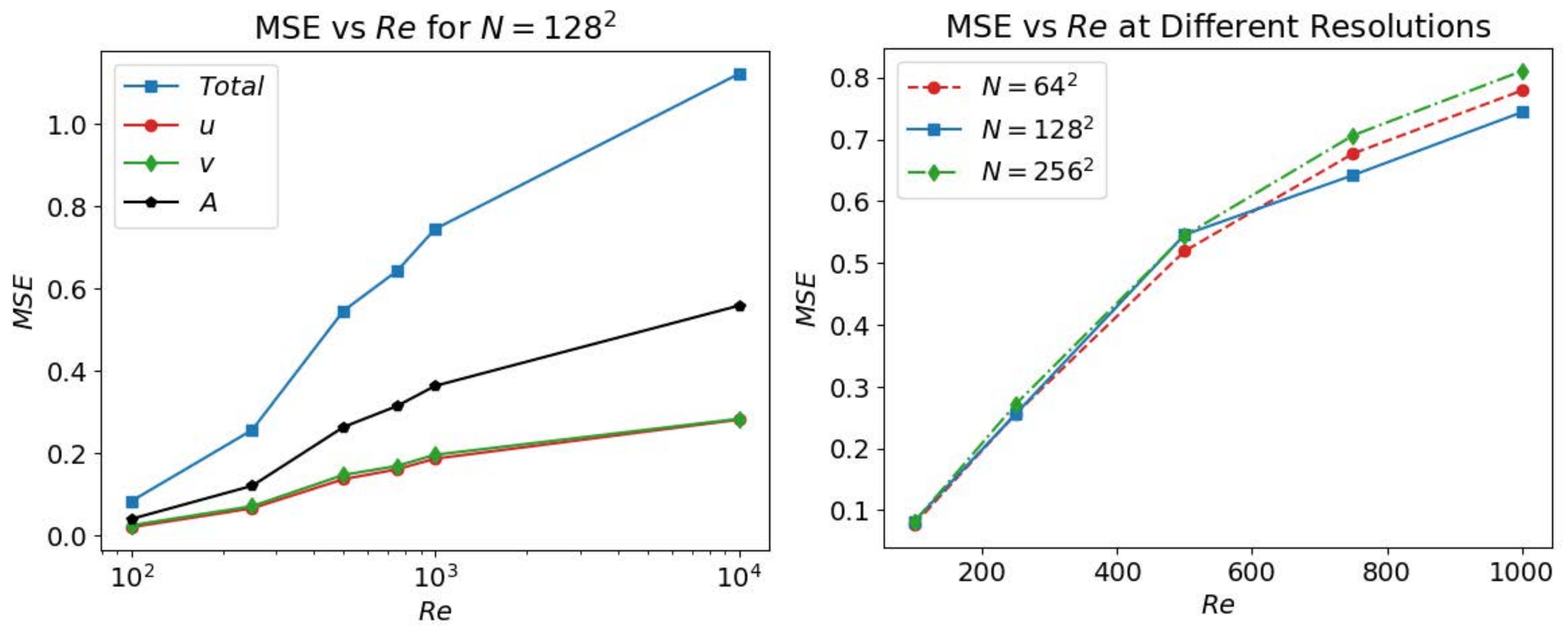}
    \caption{\textbf{Left panel} MSE vs Reynolds number $Re$ at a resolution of $N=128^2$.  We include the total MSE and the MSE for each of the fields the model is trying to reconstruct--velocity in the $x$ direction $u$, the velocity in the $y$ direction $v$, and magnetic vector potential $A$. \textbf{Right panel} Total MSE vs Reynolds number $Re$ for grids of size \(N=\{64^2,128^2, 256^2\}\).  We compare these resolutions for $Re=\{100, 250, 500, 750, 1000\}$.}
    \label{ch6:fig:loss_vs_Re}
\end{figure*}

To begin with, in the left panel of 
Figure~\ref{ch6:fig:loss_vs_Re} we present how the 
data loss MSE varies with $Re$ for \(N=128^2\). Therein we 
see the loss of each field \((u, v, A)\) as well as the 
total loss as a function of $Re$. We observe that the two velocity fields $u$ and $v$ possess around the same MSE value as one would expect.  In contrast, the MSE for the magnetic potential field $A$ is 
larger than that of the velocity fields.  Moreover, the MSE value of the $A$ field increases faster compared to those of the velocity fields $u$ and $v$. 
In the right panel of Figure~\ref{ch6:fig:loss_vs_Re} we 
see the total MSE as a function of $Re$ for three different 
resolutions, \(N=\{64^2,128^2, 256^2\}\). 
If we combine these two sets of results, we realize that 
our AI surrogates can produce reliable 
MHD simulations for \(Re \leq 250\), and that the main 
factor that degrades the accuracy of our AI surrogates is 
the ability to capture detailed features of the vector potential 
$A$ in turbulent flows. We also notice that the resolution of the 
training data does not have a major impact in the accuracy of our 
AI surrogates.

\subsection{PDE Weight}
\label{ch6:sec:pde_weight}
\begin{figure*}[htb!]
    \centerline{
    \includegraphics[width=\textwidth]{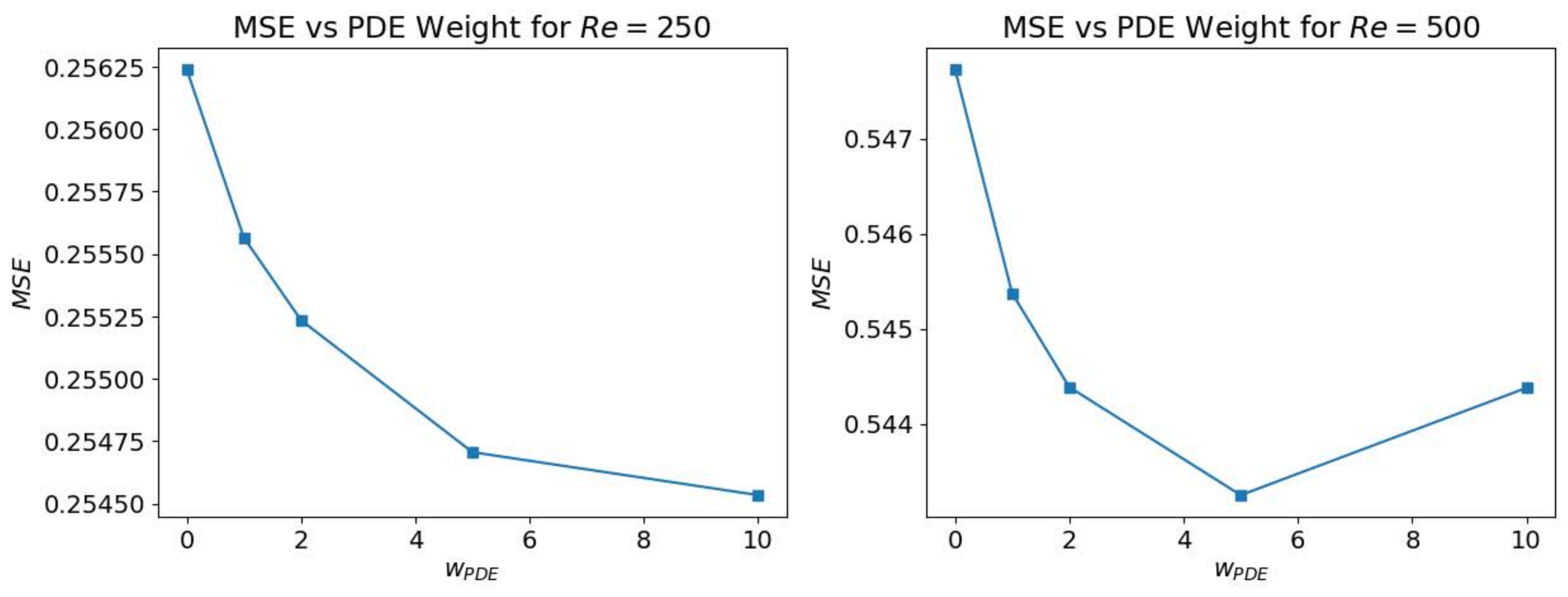}
    }
    \caption{\textbf{MSE vs PDE weight} Impact of PDE weight, $w_{\textrm{PDE}}$, on the performance of AI surrogates in 
    terms of total MSE for $Re=250$ (left panel). 
    and $Re=500$ (right panel) simulations.  We show 
    results for $w_{\textrm{PDE}}$ values of $0$, $1$, $2$, $5$, and $10$.}
    \label{ch6:fig:loss_pde_weight}
\end{figure*}

Figure~\ref{ch6:fig:loss_pde_weight} illustrates how the MSE changes with PDE weight, $w_{\textrm{PDE}}$, for $Re=250$ and $Re=500$.  We observe that the MSE improves as we increase $w_{\textrm{PDE}}$.  The only exception of $Re=500$ and $w_{\textrm{PDE}}=10$ which increases in MSE compared to $w_{\textrm{PDE}}=5$, but still less than the other $w_{\textrm{PDE}}$ values at $Re=500$.  However, we should observe that while including violations of the PDE into the loss function improves the result, their contribution is 
only marginal.

\subsection{Timesteps}
\label{ch6:sec:pde_tsep}
\begin{figure*}[htbp!]
    \centering
    \includegraphics[width=0.49\textwidth]{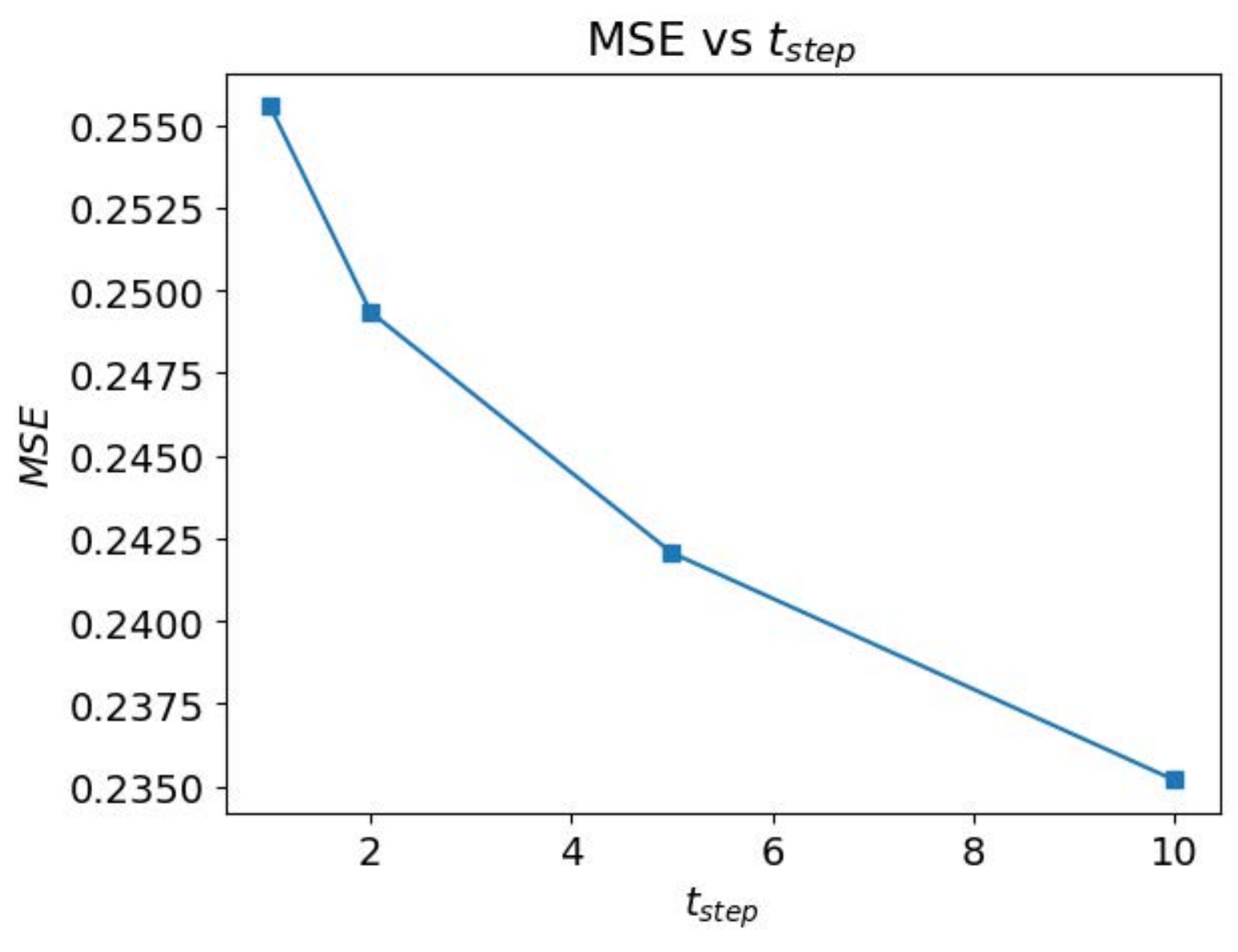}
    \caption{\textbf{MSE vs Timestep} Total MSE vs timestep, $t_{\textrm{step}}$, for values $t_{step}=\{1, 2, 5, 10\}$.}
    \label{ch6:fig:loss_tstep}
\end{figure*}

Figure~\ref{ch6:fig:loss_tstep} illustrates, for the $Re=250$ model, how the total MSE varies with the number of 
timesteps, $t_{\textrm{step}}$. We observe that the fewer timesteps the model is required to output data for (higher $t_{step}$), the better the performance of the model---though this improvement is marginal.

\subsection{Pair-wise comparison of traditional and AI-driven MHD simulations}
\label{ch6:sec:pair_wise}
\begin{figure*}[!htbp]
    \centering
    \includegraphics[width=\textwidth]{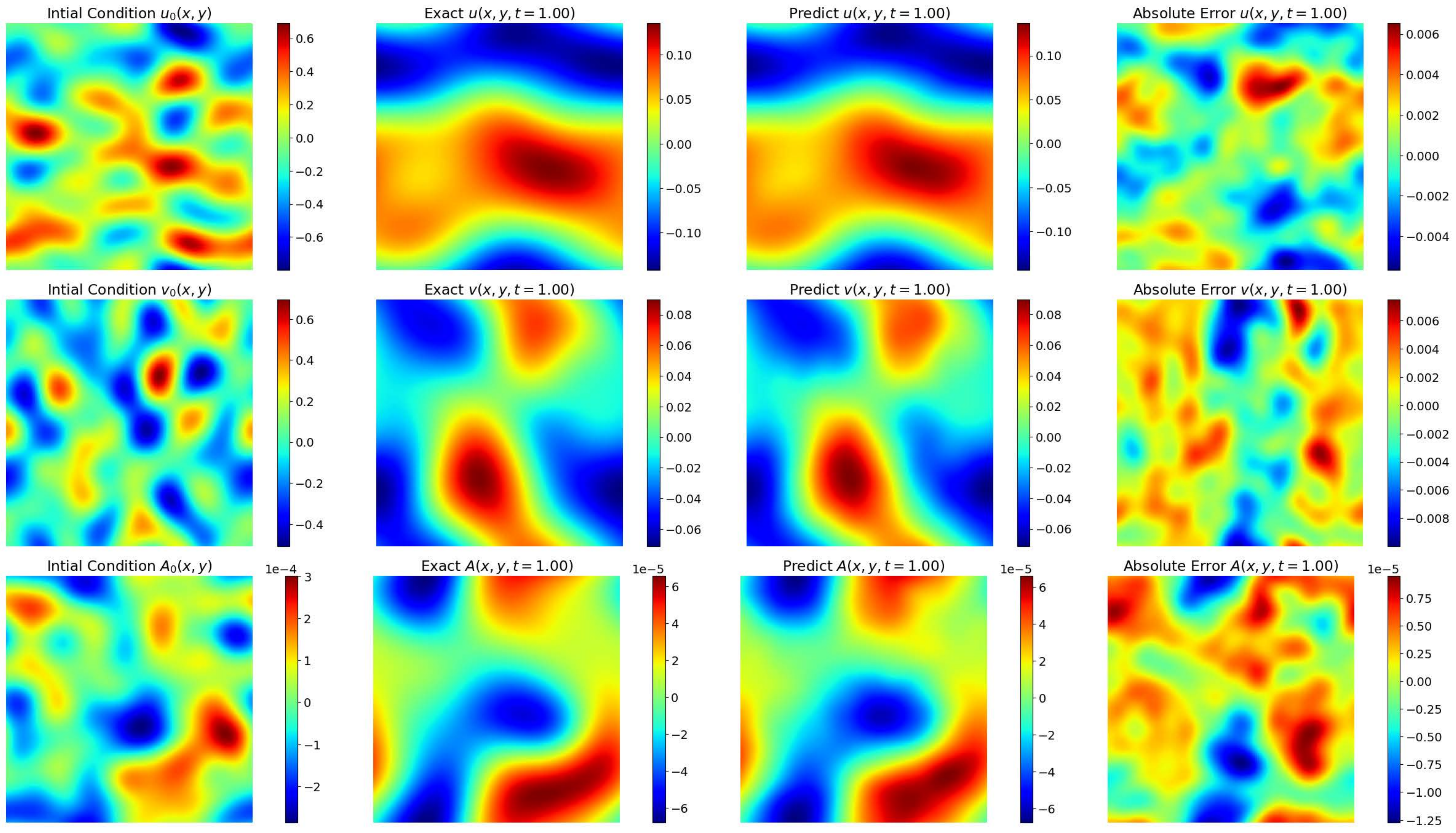}
    \caption{\textbf{$\mathbf{Re=100}$ MHD Simulations} 
    Results of the PINO model on sample test data from the 
    $Re=100$ simulations. The rows correspond to the fields 
    of interest with the velocity in the x direction $u$, the 
    velocity in the y direction $v$, and the magnetic potential 
    $A$ being the quantities featured in the top, middle, bottom 
    rows, respectively. The far left column depicts the 
    initial condition given to the PINO model. The center 
    left column shows the ground truth at time $t=1$.  In the 
    center right column, we present the PINO predictions at 
    time $t=1$.  The far right column error between the ground 
    truth and the PINO predictions at $t=1$.}
    \label{ch6:fig:Re100_tfno}
\end{figure*}

\begin{figure*}[!htbp]
    \centering
    \includegraphics[width=\textwidth]{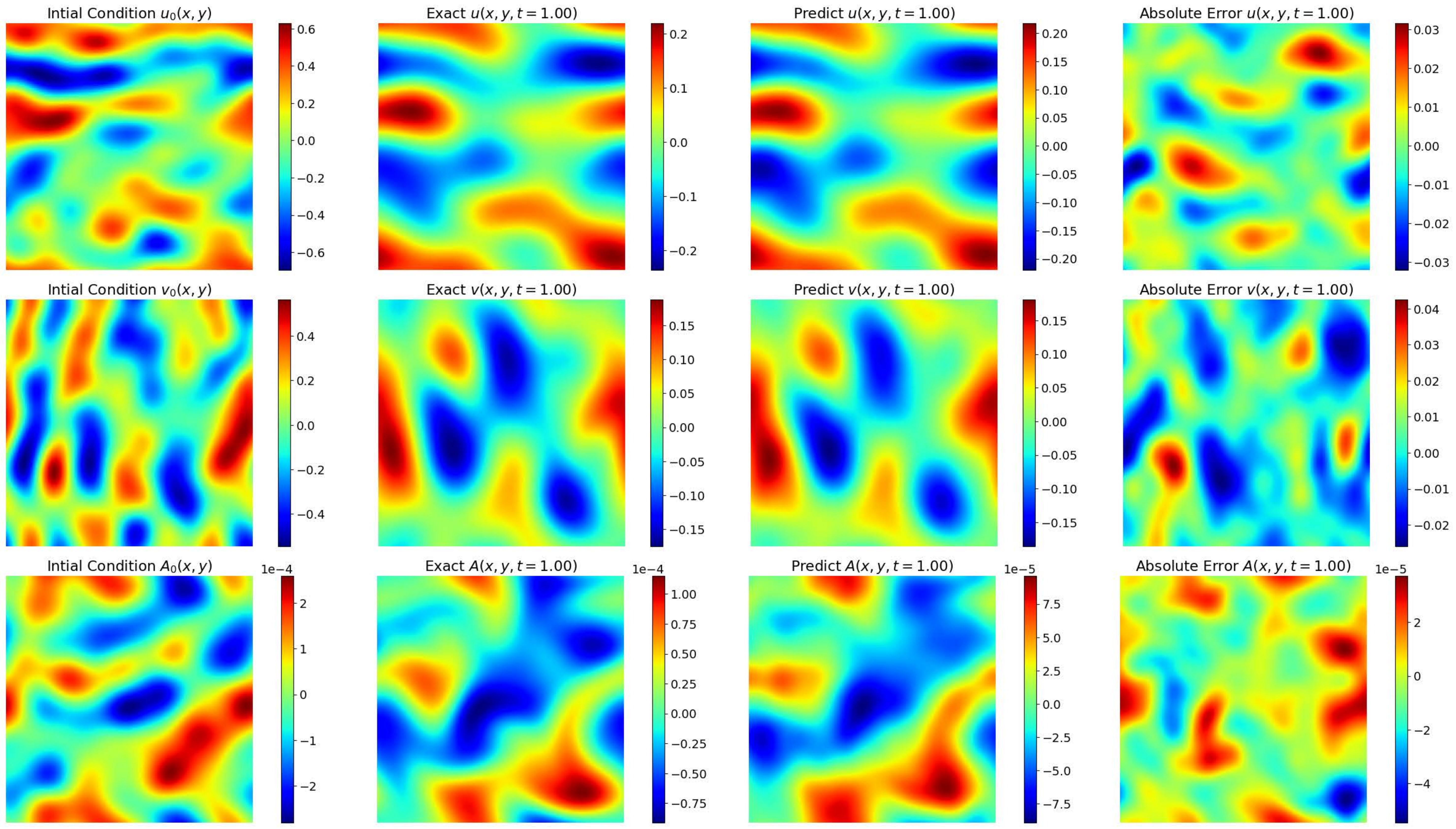}
    \caption{\textbf{$\mathbf{Re=250}$ MHD Simulations} 
    Same as Figure~\ref{ch6:fig:Re100_tfno} but now for a test 
    set from the $Re=250$ simulations.}
    \label{ch6:fig:Re250_tfno}
\end{figure*}

\begin{figure*}[!htbp]
    \centering
    \includegraphics[width=\textwidth]{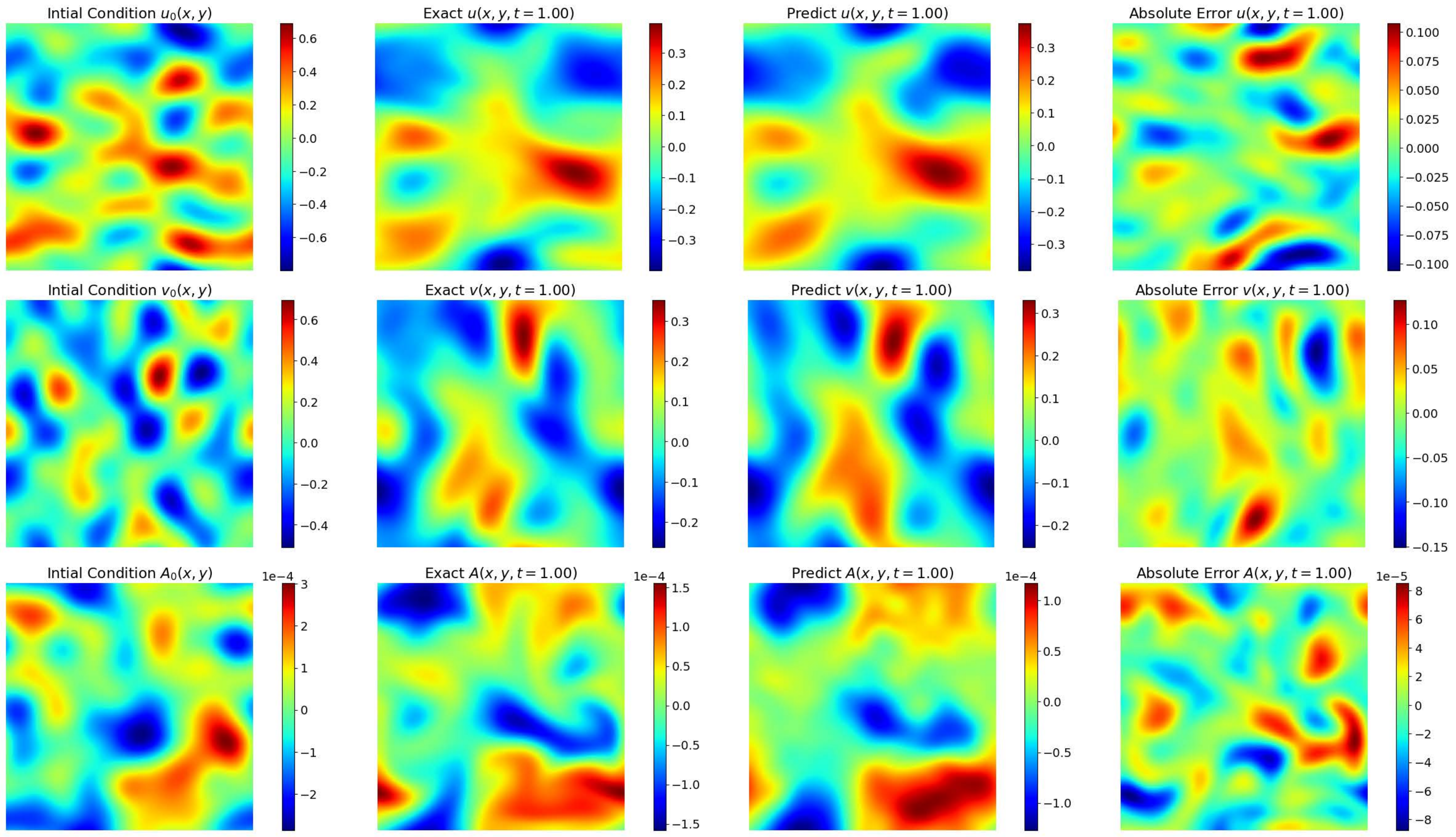}
    \caption{\textbf{$\mathbf{Re=500}$ MHD Simulations} 
    Same as Figure~\ref{ch6:fig:Re100_tfno} but now for a test set 
    from the $Re=500$ simulations.}
    \label{ch6:fig:Re500_tfno}
\end{figure*}

\begin{figure*}[!htbp]
    \centering
    \includegraphics[width=\textwidth]{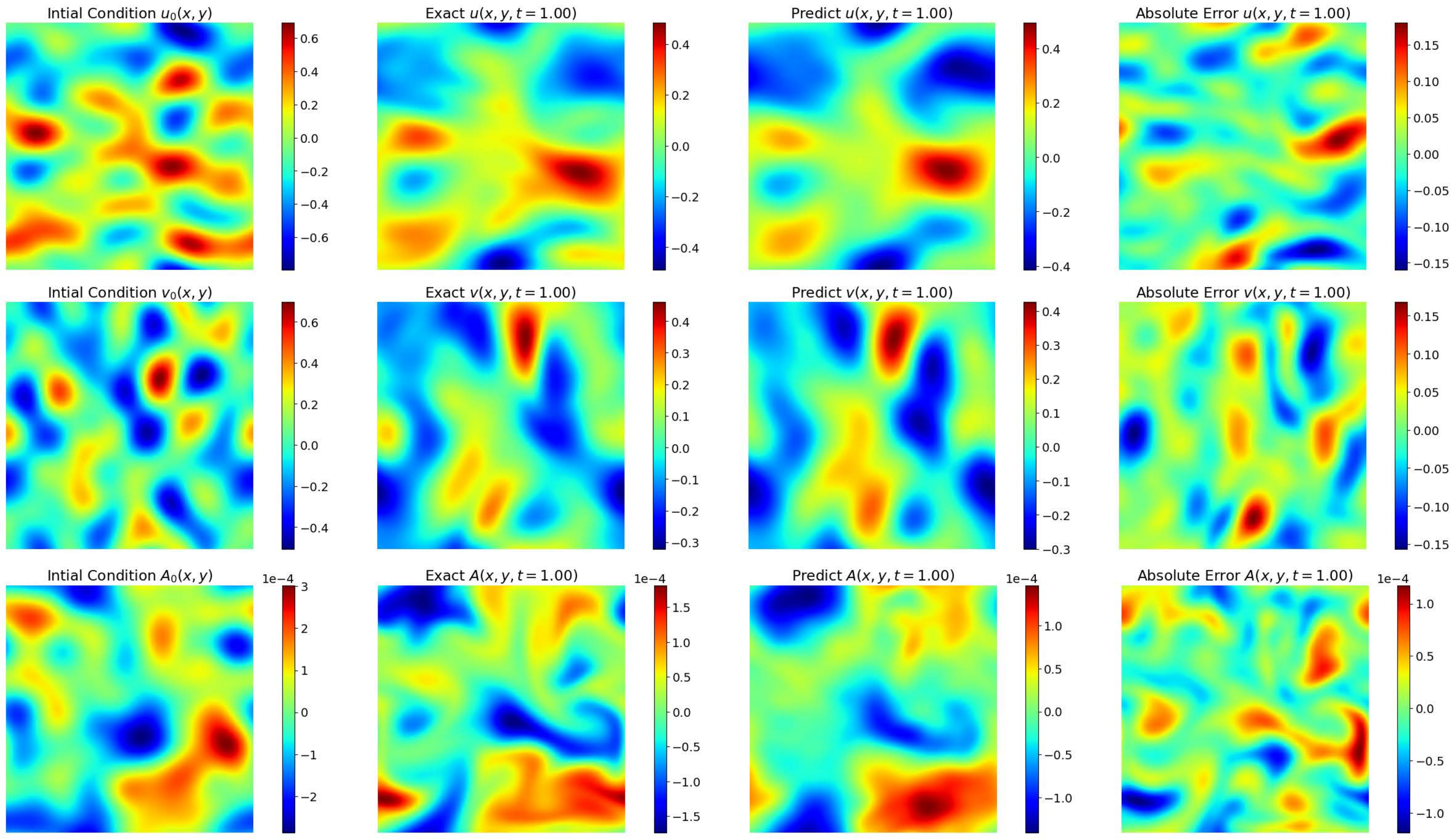}
    \caption{\textbf{$\mathbf{Re=750}$ MHD Simulations} 
    Same as Figure~\ref{ch6:fig:Re100_tfno} but now for a test set 
    from the $Re=750$ simulations.}
    \label{ch6:fig:Re750_tfno}
\end{figure*}

\begin{figure*}[!htbp]
    \centering
    \includegraphics[width=\textwidth]{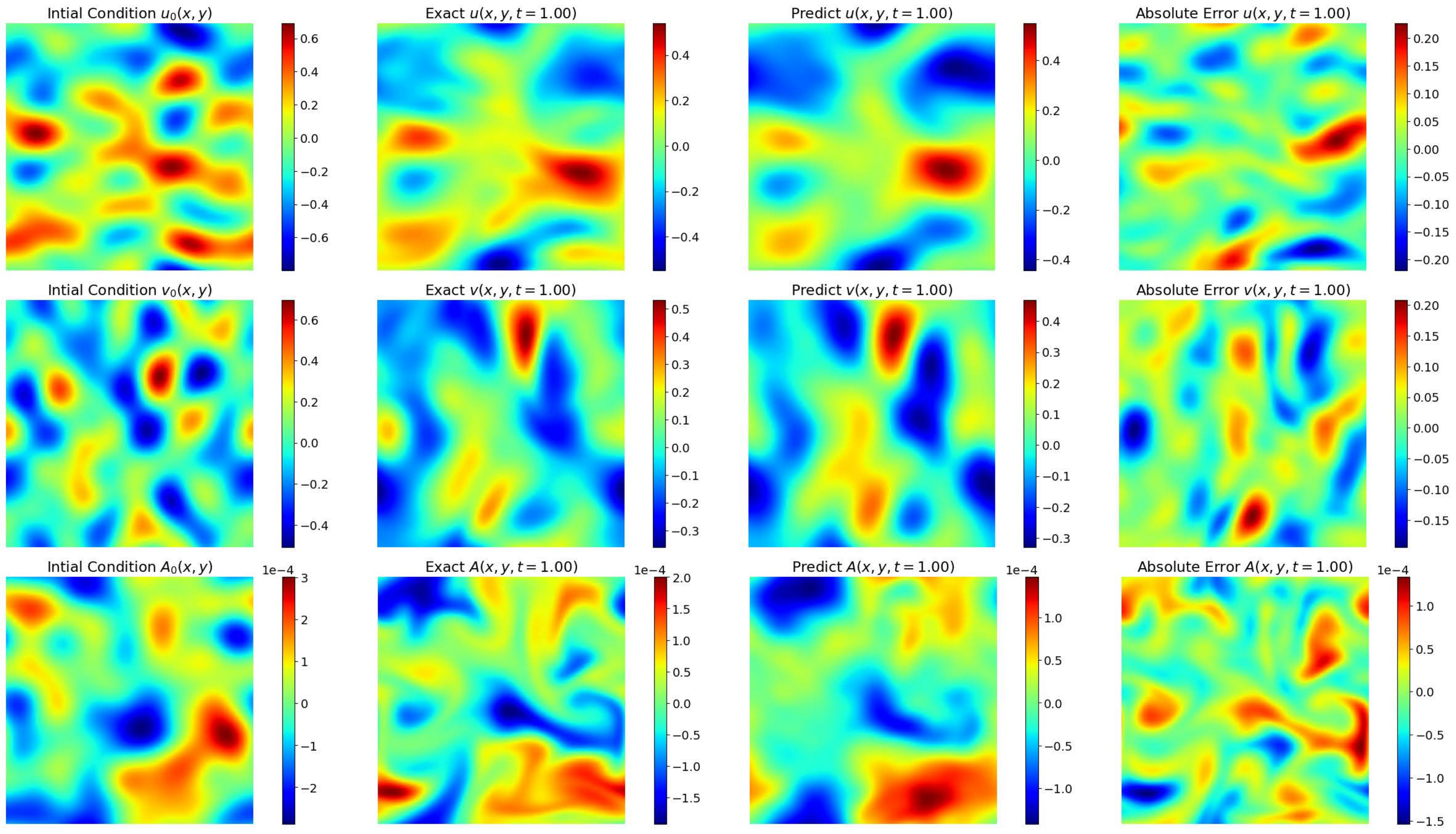}
    \caption{\textbf{$\mathbf{Re=1,000}$ MHD Simulations} 
    Same as Figure~\ref{ch6:fig:Re100_tfno} but now for a test set 
    from the $Re=1,000$ simulations.}
    \label{ch6:fig:Re1000_tfno}
\end{figure*}

\begin{figure*}[!htbp]
    \centering
    \includegraphics[width=\textwidth]{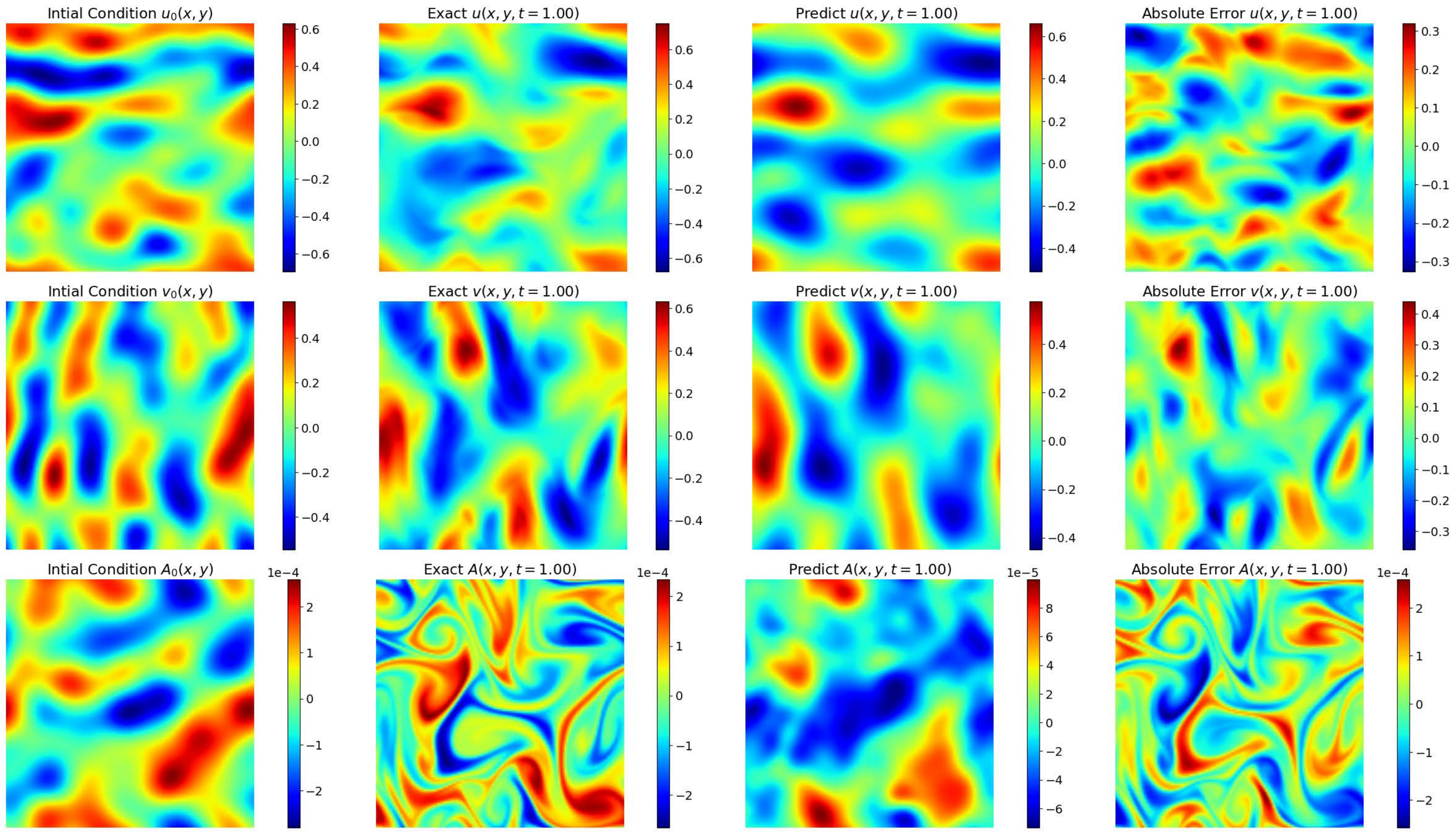}
    \caption{\textbf{$\mathbf{Re=10,000}$ MHD Simulations} 
    Same as Figure~\ref{ch6:fig:Re100_tfno} but now for a test set 
    from the $Re=10,000$ simulations.}
    \label{ch6:fig:Re10000_tfno}
\end{figure*}

We now compare traditional MHD simulations with predictions 
from AI surrogates assuming a grid of size \(N=128^2\), and 
$t_{step}=1$. In 
Figure~\ref{ch6:fig:Re100_tfno}-\ref{ch6:fig:Re10000_tfno}
we present snapshots of the systems' evolution at 
time $t=1$ to show the largest discrepancy 
between ground truth values (traditional MHD simulations) 
and AI predictions.
These results illustrate initial conditions, 
targets, prediction, and MSEs between target and 
AI predictions. At a glance, we observe that 
AI models appear
to possess the ability to resolve large scale features, 
but struggle to resolve cases with small scale structure.
This small scale structure arise at higher $Re$ and is observed
most strongly in the $A$ field. We summarize below the main findings of these 
studies using Table~\ref{ch6:tab:pino_results}, Figure~\ref{ch6:fig:loss_vs_Re} and Figures~\ref{ch6:fig:Re100_tfno}-\ref{ch6:fig:Re10000_tfno}:

\begin{itemize}
    \item $Re=100$. Figure~\ref{ch6:fig:Re100_tfno} 
    shows that PINO models provide an accurate 
    description of these MHD simulations. Quantitatively 
    and quantitatively PINOs can resolve the dynamics 
    of the velocity field, \((u,v)\), and the vector 
    potential, $A$. We also notice that the 
    largest discrepancy between AI predictions 
    and traditional MHD simulations at \(t=1\) is 
    \(\leq 4\%\) for each of the fields.
    \item $Re=250$. Figure~\ref{ch6:fig:Re250_tfno} 
    shows that PINOs provide a reliable 
    description of the dynamics 
    of these simulations. The velocity 
    field and vector potential potential are accurately described, with MSEs \(\leq 7\%\) and \(\leq 10\%\), 
    respectively.
    \item $Re=500$. Figure~\ref{ch6:fig:Re500_tfno} 
    shows that PINOs capture well large scale features of these 
    simulations. In particular, the velocity field 
    can be recovered with MSE \(\leq 14\%\). On 
    the other hand, detailed features of the vector potential 
    are not completely resolved, and the MSE is \(\leq 26\%\). 
    \item $Re=750$. Figure~\ref{ch6:fig:Re750_tfno} 
    shows that PINO MHD simulations can resolve large scale 
    features of the velocity field, with MSE \(\leq 16\%\), 
    which is similar to simulations 
    with $Re=500$. Similarly, large scale features of the 
    vector potential are well described. However, as these systems become 
    more turbulent, small scale features of the vector potential 
    are not captured by PINOs, and report an increase in MSE, i.e., \(\leq 31\%\).
    \item $Re=1,000$. Figure~\ref{ch6:fig:Re1000_tfno} presents a similar story to the two previous cases. Large 
    scale structure is well described, and the MSE for 
    the velocity field is \(\leq 18\%\). However, it is 
    difficult to capture the small scale features of the 
    magnetic field, which now evolve in a rather complex 
    manner for these turbulent systems.
    \item $Re=10,000$. Figure~\ref{ch6:fig:Re10000_tfno} indicates that, for 
    very turbulent MHD systems, PINOs can 
    only reproduce some of the large scale features of 
    the flow, but struggle to reproduce many of the detailed features. One particular feature that the PINO misses 
    is the small scale fluctuations in the velocity fields 
    $u$ and $v$ that likely result from strong magnetic fields 
    in those areas. In other words, the PINO cannnot resolve the effect of the magnetic field on the fluid motion. For the magnetic potential $A$, the PINO model appears to miss most of the important features at first glance. If we look more closely, we observe that the PINO appears, to some extent, to reproduce the mean field value over some very large regions. However, this comes at the cost of missing any sort of interesting details in the magnetic field.
\end{itemize}

In summary, Figures~\ref{ch6:fig:Re100_tfno}-\ref{ch6:fig:Re10000_tfno} 
show that our PINO models can successfully reproduce 
large scale features of MHD simulations, but had 
difficulty resolving detailed features for large Reynolds 
numbers, especially for 
the magnetic potential, which is known to store its energy 
at higher wavenumbers. Thus, in the following 
section we explored the ability of PINOs to 
learn the right data features associated to low 
and high wavenumbers as the simulation evolves in time.

\subsection{Spectra Results}
\label{ch6:sec:spectra_results}
\begin{figure*}[htbp]
    \centering
    \includegraphics[width=\textwidth]{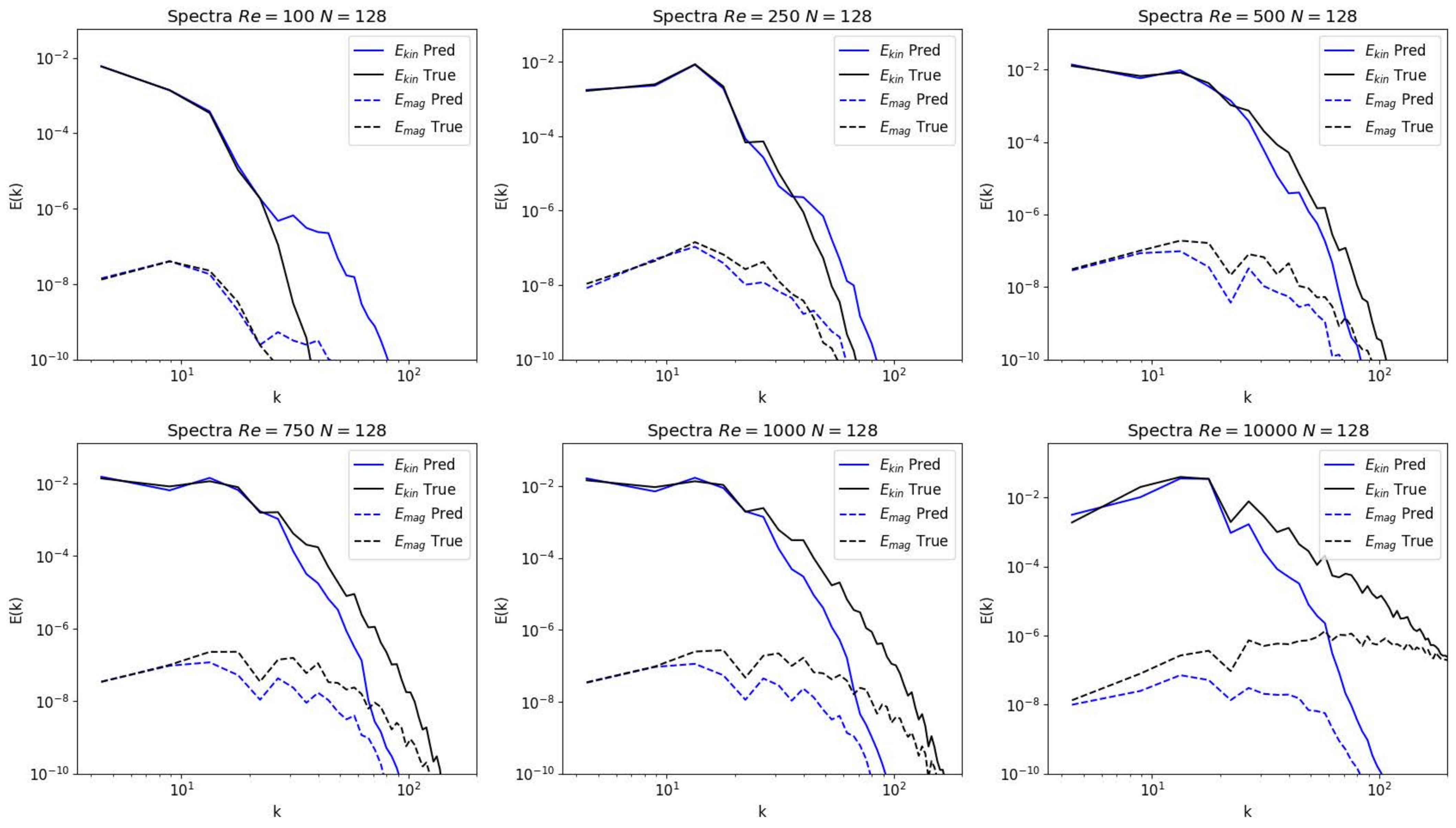}
    \caption{\textbf{Spectra} The panels show the kinetic and 
    magnetic spectra for each PINO model and simulation at 
    time $t=1$.  We present the kinetic energy spectra as 
    solid lines and the magnetic energy spectra as dashed lines. 
    The ground truth simulations are illustrated in black while 
    the PINO predictions are in blue. The top row presents 
    from left to right the $Re=100$, $Re=250$, and $Re=500$. 
    The bottom row contains 
    the $Re=750$, $Re=1,000$, and $Re=10,000$ cases.}
    \label{ch6:fig:spectra}
\end{figure*}

We analyzed the magnetic 
and kinetic energy spectra of the simulations 
in Figure~\ref{ch6:fig:spectra}. Therein we observe that for both 
the kinetic and magnetic energy spectra, the PINO models 
performed well at low wavenumbers $k$.  However, the PINOs 
were not able to accurately reproduce the energy spectra at 
high wavenumbers. Interestingly, at low $Re$, the PINOs tended 
to overshoot the ground truth energies at high wavenumbers. 
While at high $Re$, the PINOs tended to undershoot the ground 
truth energies at high wavenumbers.

Moreover, large Reynolds number simulations usually store 
more energy at higher wavenumbers. This is especially true for 
the magnetic energy, which tends to peak at later wavenumbers 
in higher $Re$ simulations. Therefore, the spectra plots in 
Figure~\ref{ch6:fig:spectra} indicate that the difficulties the 
PINOs had in reproducing the simulations at high $Re$ and 
the magnetic potential field $A$ may stem from their 
relatively poor performance at high wavenumbers. 
Therefore, future work should focus on developing 
methods that enable PINOs to capture data features 
contained at high wavenumbers.

\section{Conclusions}
\label{ch6:sec:conclusions}

In this work we presented the first application of 
PINOs to produce 2D incompressible MHD 
simulations for a variety of 
initial conditions and Reynolds numbers. We 
demonstrated how to incorporate physics principles, 
and suitable gauge conditions, for the training and 
optimization of our AI surrogates.
Once fully trained, we found that our PINO models 
were able to accurately describe 
MHD simulations with Reynolds numbers $Re\leq 250$ 
throughout the entire evolution of 
the system, i.e., \(t\in[0,1]\). For other systems with 
larger Reynolds numbers, we found that PINOs provide 
a reliable description of the system at earlier 
times, but as the system evolves, PINO simulations 
gradually degrade in accuracy. We first noticed 
this behaviour in the evolution of 
the magnetic potential, $A$, which we used to evolve the magnetic 
fields, for MHD simulations with $Re> 500$.  
We explored this issue in detail, and found that 
this issue may likely stem from the PINOs' difficulty 
in learning physics contained at high wavenumbers.

We suggest that future work should focus on optimizing 
neural operators at these high wavenumbers. One suggestion would 
be to increase the number of Fourier modes used by our tFNO 
backend. We were memory limited and were restricted to 8 
Fourier modes to store the model in GPU memory. 
Recent work involving FNOs for hydrodynamic turbulence 
modeling have had success with using 20 Fourier 
modes~\cite{Li2022:FNO_LES}. One should be cautious, however, 
since using too many Fourier modes may also introduce 
numerical noise and complicate the resolution of 
small scale features in the data. We also 
suggest to use emergent methods to train PINOs  
using high resolution datasets which, while difficult to 
fit in a state-of-the-art GPU, may be readily 
used in AI-accelerator machines, as those housed 
in multiple supercomputing centers in the US and 
elsewhere, e.g., the 
\href{https://www.alcf.anl.gov/alcf-ai-testbed}{Argonne Leadership 
Computing Facility AI-Testbed.}

Another suggestion is to develop methods that enable 
PINOs to learn the turbulence statistics of the MHD simulation 
in addition to the flow. For example, one could incorporate 
the kinetic and magnetic energy spectra into the loss function and 
weigh in the contribution of the high wavenumber portions 
of the spectrum appropriately. Through this approach, 
PINOs may may predict the flow with the correct 
turbulent characteristics even if they are unable to accurately 
model the details of the flow. It is also advisable to go beyond the use of physics-informed 
loss functions, which only provide static optimization at present. 
A more suitable approach for these types of complex systems would 
entail coupling physics-informed loss functions with online or 
reinforcement learning that dynamically steer the 
AI surrogate to the right answer during the 
optimization procedure as new data features, such as 
magnetic fields, arise in the simulation data. 
We expect that the work we have presented here, 
in terms of data generators, AI surrogates, and 
optimization approaches for complex systems,
may provide a stepping stone to other AI practitioners who 
are developing novel methods to model complex systems, 
such as turbulent MHD simulations.

\section*{Acknowledgments}
\noindent  This material is based upon work supported by 
Laboratory Directed Research and Development (LDRD) funding 
from Argonne National Laboratory, provided by the Director, 
Office of Science, of the U.S. Department of Energy under 
Contract No. DE-AC02-06CH11357. This research used resources 
of the Argonne Leadership Computing Facility, which is a 
DOE Office of Science User Facility supported under 
Contract DE-AC02-06CH11357. S.R. and E.A.H. gratefully 
acknowledge National Science Foundation award OAC-1931561. 
This work used the Extreme Science and Engineering 
Discovery Environment (XSEDE), which is supported by 
National Science Foundation grant number ACI-1548562. 
This work used the Extreme Science and Engineering 
Discovery Environment (XSEDE) Bridges-2 at the 
Pittsburgh Supercomputing Center through allocation 
TG-PHY160053. This research used the Delta advanced 
computing and data resource which is supported by the 
National Science Foundation (award OAC-2005572) and the 
State of Illinois. Delta is a joint effort of the 
University of Illinois Urbana-Champaign and its National 
Center for Supercomputing Applications. 

\section*{References}
\bibliography{references}

\providecommand{\newblock}{}
\begin{thebibliography}{10}
\expandafter\ifx\csname url\endcsname\relax
  \def\url#1{{\tt #1}}\fi
\expandafter\ifx\csname urlprefix\endcsname\relax\def\urlprefix{URL }\fi
\providecommand{\eprint}[2][]{\url{#2}}

\bibitem{Beresnyak2019:mhd_turbulence_review}
Beresnyak A 2019 {\em Living Reviews in Computational Astrophysics\/} {\bf 5} 2
  ISSN 2365-0524 \urlprefix\url{https://doi.org/10.1007/s41115-019-0005-8}

\bibitem{schekochihin2022:mhd_turbulence_review}
Schekochihin A~A 2022 {\em Journal of Plasma Physics\/} {\bf 88} 155880501

\bibitem{Pouquet2019:mhd_turbulence_review}
Pouquet A, Rosenberg D, Stawarz J and Marino R 2019 {\em Earth and Space
  Science\/} {\bf 6} 351--369 (\textit{Preprint}
  \eprint{https://agupubs.onlinelibrary.wiley.com/doi/pdf/10.1029/2018EA000432})
  \urlprefix\url{https://agupubs.onlinelibrary.wiley.com/doi/abs/10.1029/2018EA000432}

\bibitem{Kuichi2015:BNS_KHI_highres}
Kiuchi K, Cerd\'a-Dur\'an P, Kyutoku K, Sekiguchi Y and Shibata M 2015 {\em
  Phys. Rev. D\/} {\bf 92}(12) 124034
  \urlprefix\url{https://link.aps.org/doi/10.1103/PhysRevD.92.124034}

\bibitem{Beresnyak2012:dynamo}
Beresnyak A 2012 {\em Phys. Rev. Lett.\/} {\bf 108}(3) 035002
  \urlprefix\url{https://link.aps.org/doi/10.1103/PhysRevLett.108.035002}

\bibitem{Beresnyak2015}
Beresnyak A and Lazarian A 2015 {\em MHD Turbulence, Turbulent Dynamo and
  Applications\/} (Berlin, Heidelberg: Springer Berlin Heidelberg) pp 163--226
  ISBN 978-3-662-44625-6
  \urlprefix\url{https://doi.org/10.1007/978-3-662-44625-6_8}

\bibitem{Grete2017}
{Grete} P 2017 {\em {Large eddy simulations of compressible magnetohydrodynamic
  turbulence}\/} Ph.D. thesis Max-Planck-Institut f{\"u}r Sonnensystemforschung

\bibitem{Grete2015}
Grete P, Vlaykov D~G, Schmidt W, Schleicher D~R~G and Federrath C 2015 {\em New
  Journal of Physics\/} {\bf 17} 023070
  \urlprefix\url{https://doi.org/10.1088%2F1367-2630%2F17%2F2%2F023070}

\bibitem{Grete2016}
Grete P, Vlaykov D~G, Schmidt W and Schleicher D~R~G 2016 {\em Physics of
  Plasmas\/} {\bf 23} 062317 (\textit{Preprint}
  \eprint{https://doi.org/10.1063/1.4954304})
  \urlprefix\url{https://doi.org/10.1063/1.4954304}

\bibitem{Vlaykov2016}
Vlaykov D~G, Grete P, Schmidt W and Schleicher D~R~G 2016 {\em Physics of
  Plasmas\/} {\bf 23} 062316 (\textit{Preprint}
  \eprint{https://doi.org/10.1063/1.4954303})
  \urlprefix\url{https://doi.org/10.1063/1.4954303}

\bibitem{Grete2017a}
Grete P, Vlaykov D~G, Schmidt W and Schleicher D~R~G 2017 {\em Phys. Rev. E\/}
  {\bf 95}(3) 033206
  \urlprefix\url{https://link.aps.org/doi/10.1103/PhysRevE.95.033206}

\bibitem{Grete2017b}
Grete P, O'Shea B~W, Beckwith K, Schmidt W and Christlieb A 2017 {\em Physics
  of Plasmas\/} {\bf 24} 092311 (\textit{Preprint}
  \eprint{https://doi.org/10.1063/1.4990613})
  \urlprefix\url{https://doi.org/10.1063/1.4990613}

\bibitem{Kessar2016}
Kessar M, Balarac G and Plunian F 2016 {\em Physics of Plasmas\/} {\bf 23}
  102305 (\textit{Preprint} \eprint{https://doi.org/10.1063/1.4964782})
  \urlprefix\url{https://doi.org/10.1063/1.4964782}

\bibitem{Aguilera-Miret2022:LES_BNS}
Aguilera-Miret R, Viganò D and Palenzuela C 2022 {\em The Astrophysical
  Journal Letters\/} {\bf 926} L31
  \urlprefix\url{https://dx.doi.org/10.3847/2041-8213/ac50a7}

\bibitem{Palenzuela2022:LES_BNS}
Palenzuela C, Aguilera-Miret R, Carrasco F, Ciolfi R, Kalinani J~V, Kastaun W,
  Mi\~nano B and Vigan\`o D 2022 {\em Phys. Rev. D\/} {\bf 106}(2) 023013
  \urlprefix\url{https://link.aps.org/doi/10.1103/PhysRevD.106.023013}

\bibitem{Vigano2019:MHD_LES}
Viganò D, Aguilera-Miret R and Palenzuela C 2019 {\em Physics of Fluids\/}
  {\bf 31} 105102 \urlprefix\url{https://doi.org/10.1063/1.5121546}

\bibitem{Carrasco2020:MHD_LES_SR}
Carrasco F, Vigan\`o D and Palenzuela C 2020 {\em Phys. Rev. D\/} {\bf 101}(6)
  063003 \urlprefix\url{https://link.aps.org/doi/10.1103/PhysRevD.101.063003}

\bibitem{Vigano2020:MHD_LES_grad_GR}
Vigan\`o D, Aguilera-Miret R, Carrasco F, Mi\~nano B and Palenzuela C 2020 {\em
  Phys. Rev. D\/} {\bf 101}(12) 123019
  \urlprefix\url{https://link.aps.org/doi/10.1103/PhysRevD.101.123019}

\bibitem{Radice2020:BNS_SGS}
Radice D 2020 {\em Symmetry\/} {\bf 12} ISSN 2073-8994
  \urlprefix\url{https://www.mdpi.com/2073-8994/12/8/1249}

\bibitem{Rosofsky2020a:SGS_MHD_2D}
Rosofsky S~G and Huerta E~A 2020 {\em Phys. Rev. D\/} {\bf 101}(8) 084024
  \urlprefix\url{https://link.aps.org/doi/10.1103/PhysRevD.101.084024}

\bibitem{Karpov2022:supernova_ANN_LES}
Karpov P~I, Huang C, Sitdikov I, Fryer C~L, Woosley S and Pilania G 2022 {\em
  The Astrophysical Journal\/} {\bf 940} 26
  \urlprefix\url{https://dx.doi.org/10.3847/1538-4357/ac88cc}

\bibitem{kovachki2021neuraloperator}
{Kovachki} N, {Li} Z, {Liu} B, {Azizzadenesheli} K, {Bhattacharya} K, {Stuart}
  A and {Anandkumar} A 2021 {\em arXiv e-prints\/} arXiv:2108.08481
  (\textit{Preprint} \eprint{2108.08481})

\bibitem{Peng2022:FNO_3D_turbulence}
Peng W, Yuan Z, Li Z and Wang J 2022 Linear attention coupled fourier neural
  operator for simulation of three-dimensional turbulence
  \urlprefix\url{https://arxiv.org/abs/2210.04259}

\bibitem{Peng2022:FNO_attention}
Peng W, Yuan Z and Wang J 2022 {\em Physics of Fluids\/} {\bf 34} 025111
  (\textit{Preprint} \eprint{https://doi.org/10.1063/5.0079302})
  \urlprefix\url{https://doi.org/10.1063/5.0079302}

\bibitem{Li2022:FNO_LES}
Li Z, Peng W, Yuan Z and Wang J 2022 {\em Theoretical and Applied Mechanics
  Letters\/}  100389 ISSN 2095-0349
  \urlprefix\url{https://www.sciencedirect.com/science/article/pii/S2095034922000691}

\bibitem{PINO_li_anima}
{Li} Z, {Zheng} H, {Kovachki} N, {Jin} D, {Chen} H, {Liu} B, {Azizzadenesheli}
  K and {Anandkumar} A 2021 {\em arXiv e-prints\/} arXiv:2111.03794
  (\textit{Preprint} \eprint{2111.03794})

\bibitem{Rosofsky2022:PINO}
{Rosofsky} S~G, {Al Majed} H and {Huerta} E~A 2022 {\em arXiv e-prints\/}
  arXiv:2203.12634 (\textit{Preprint} \eprint{2203.12634})

\bibitem{Dedner2002}
Dedner A, Kemm F, Kröner D, Munz C~D, Schnitzer T and Wesenberg M 2002 {\em
  Journal of Computational Physics\/} {\bf 175} 645 -- 673 ISSN 0021-9991
  \urlprefix\url{http://www.sciencedirect.com/science/article/pii/S002199910196961X}

\bibitem{Mocz2014:constrained_transport}
Mocz P, Vogelsberger M and Hernquist L 2014 {\em Monthly Notices of the Royal
  Astronomical Society\/} {\bf 442} 43--55 ISSN 0035-8711 (\textit{Preprint}
  \eprint{https://academic.oup.com/mnras/article-pdf/442/1/43/4072384/stu865.pdf})
  \urlprefix\url{https://doi.org/10.1093/mnras/stu865}

\bibitem{Burns2020:Dedalus}
{Burns} K~J, {Vasil} G~M, {Oishi} J~S, {Lecoanet} D and {Brown} B~P 2020 {\em
  Physical Review Research\/} {\bf 2} 023068 (\textit{Preprint}
  \eprint{1905.10388})

\bibitem{lu2021learning}
Lu L, Jin P, Pang G, Zhang Z and Karniadakis G~E 2021 {\em Nature Machine
  Intelligence\/} {\bf 3} 218–229 ISSN 2522-5839
  \urlprefix\url{http://dx.doi.org/10.1038/s42256-021-00302-5}

\bibitem{wang2021learning}
Wang S, Wang H and Perdikaris P 2021 {\em Science Advances\/} {\bf 7} eabi8605

\bibitem{Li2020GraphNO}
Li Z, Kovachki N, Azizzadenesheli K, Liu B, Bhattacharya K, Stuart A and
  Anandkumar A 2020 Neural operator: Graph kernel network for partial
  differential equations (\textit{Preprint} \eprint{2003.03485})

\bibitem{Li2020MGNO}
Li Z, Kovachki N~B, Azizzadenesheli K, Liu B, Bhattacharya K, Stuart A~M and
  Anandkumar A 2020 {\em CoRR\/} {\bf abs/2006.09535} (\textit{Preprint}
  \eprint{2006.09535}) \urlprefix\url{https://arxiv.org/abs/2006.09535}

\bibitem{Li2021FourierNO}
Li Z, Kovachki N, Azizzadenesheli K, Liu B, Bhattacharya K, Stuart A and
  Anandkumar A 2020 Fourier neural operator for parametric partial differential
  equations (\textit{Preprint} \eprint{2010.08895})

\bibitem{Tran2021:fFNO}
{Tran} A, {Mathews} A, {Xie} L and {Ong} C~S 2021 {\em arXiv e-prints\/}
  arXiv:2111.13802 (\textit{Preprint} \eprint{2111.13802})

\bibitem{raissi2017physicsI}
Raissi M, Perdikaris P and Karniadakis G~E 2017 Physics informed deep learning
  (part i): Data-driven solutions of nonlinear partial differential equations
  (\textit{Preprint} \eprint{1711.10561})

\bibitem{raissi2017physicsII}
Raissi M, Perdikaris P and Karniadakis G~E 2017 Physics informed deep learning
  (part ii): Data-driven discovery of nonlinear partial differential equations
  (\textit{Preprint} \eprint{1711.10566})

\bibitem{raissi2019physics}
Raissi M, Perdikaris P and Karniadakis G 2019 {\em Journal of Computational
  Physics\/} {\bf 378} 686--707 ISSN 0021-9991
  \urlprefix\url{https://www.sciencedirect.com/science/article/pii/S0021999118307125}

\bibitem{Pang2019fPINNs}
Pang G, Lu L and Karniadakis G~E 2019 {\em SIAM Journal on Scientific
  Computing\/} {\bf 41} A2603--A2626 (\textit{Preprint}
  \eprint{https://doi.org/10.1137/18M1229845})
  \urlprefix\url{https://doi.org/10.1137/18M1229845}

\bibitem{lu2021deepxde}
Lu L, Meng X, Mao Z and Karniadakis G~E 2021 {\em SIAM Review\/} {\bf 63}
  208--228

\bibitem{Karniadakis2021:physics_informed_machine_learning}
Karniadakis G~E, Kevrekidis I~G, Lu L, Perdikaris P, Wang S and Yang L 2021
  {\em Nature Reviews Physics\/} {\bf 3} 422--440 ISSN 2522-5820
  \urlprefix\url{https://doi.org/10.1038/s42254-021-00314-5}

\bibitem{tensorly}
Kossaifi J, Panagakis Y, Anandkumar A and Pantic M 2019 {\em Journal of Machine
  Learning Research\/} {\bf 20} 1--6
  \urlprefix\url{http://jmlr.org/papers/v20/18-277.html}

\bibitem{Courant1928:CFL}
Courant R, Friedrichs K and Lewy H 1928 {\em Mathematische Annalen\/} {\bf 100}
  32--74 ISSN 1432-1807 \urlprefix\url{https://doi.org/10.1007/BF01448839}

\bibitem{erichson2020:randomized_CP}
Erichson N~B, Manohar K, Brunton S~L and Kutz J~N 2020 {\em Machine Learning:
  Science and Technology\/} {\bf 1} 025012

\bibitem{hendrycks2016GELU}
Hendrycks D and Gimpel K 2016 Gaussian error linear units (gelus)
  (\textit{Preprint} \eprint{1606.08415})

\bibitem{paszke2017:PyTorch}
Paszke A, Gross S, Chintala S, Chanan G, Yang E, DeVito Z, Lin Z, Desmaison A,
  Antiga L and Lerer A 2017 Automatic differentiation in pytorch {\em NIPS 2017
  Workshop on Autodiff\/}
  \urlprefix\url{https://openreview.net/forum?id=BJJsrmfCZ}

\bibitem{loshchilov2017:AdamW}
Loshchilov I and Hutter F 2017 Decoupled weight decay regularization
  (\textit{Preprint} \eprint{1711.05101})

\bibitem{wandb}
Biewald L 2020 Experiment tracking with weights and biases software available
  from wandb.com \urlprefix\url{https://www.wandb.com/}

\end{thebibliography}



\appendix
\section{Additional results for low and high resolution simulations}
\label{sec:appex}

\begin{figure*}[htbp]
    \centering
    \includegraphics[width=\textwidth]{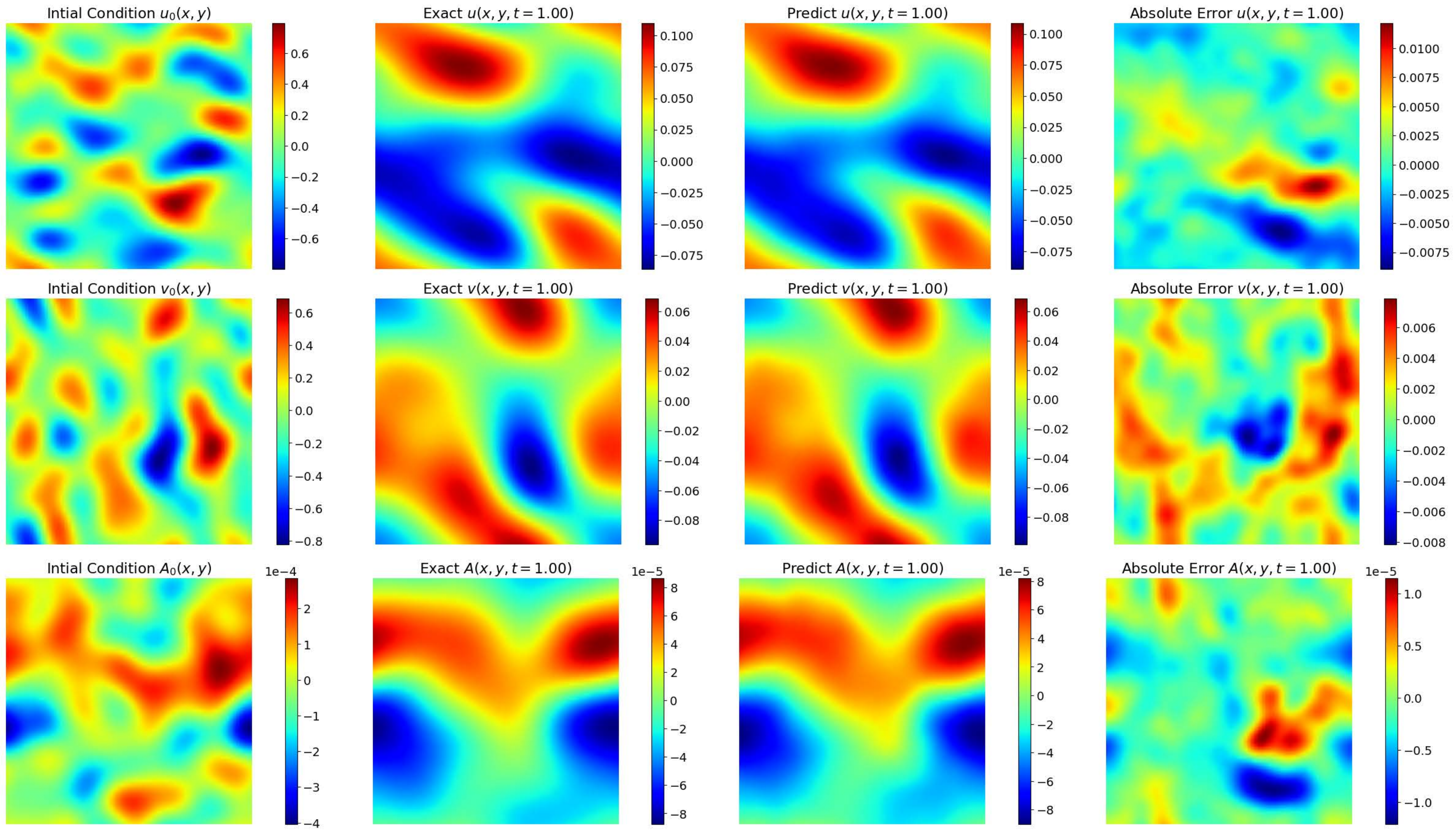}
    \caption{\textbf{$\mathbf{Re=100}$ MHD Simulations} 
    As Figure~\ref{ch6:fig:Re100_tfno} but now 
    for a test set with resolution of $N=64^2$ grid points.}
    \label{fig:Re100_r_64}
\end{figure*}

\begin{figure*}[htbp]
    \centering
    \includegraphics[width=\textwidth]{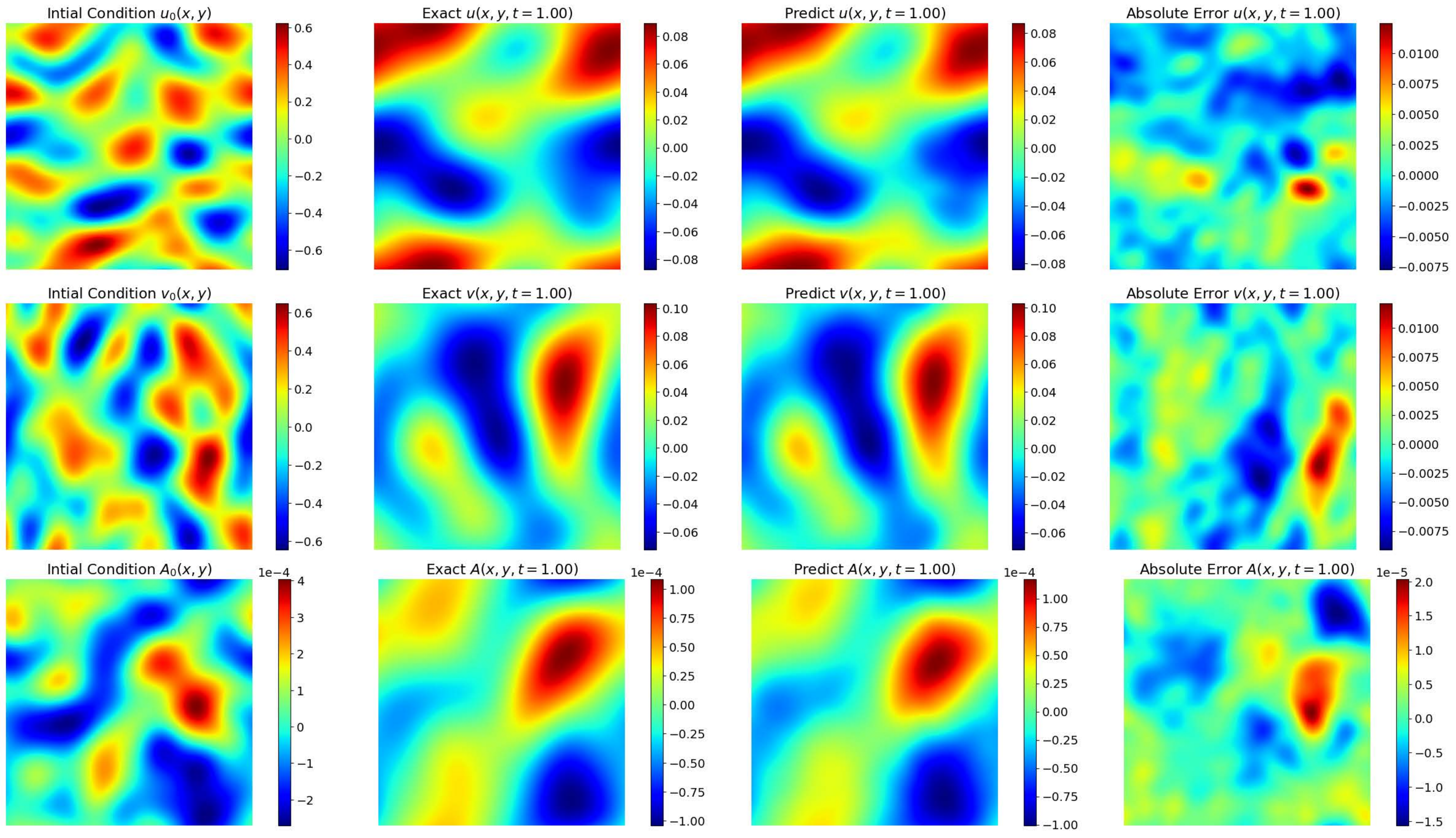}
    \caption{\textbf{$\mathbf{Re=100}$ MHD Simulations} 
    As Figure~\ref{fig:Re100_r_64} but now for a test 
    set with resolution of \(N=256^2\) grid points.}
    \label{fig:Re100_r_256}
\end{figure*}

\begin{figure*}[htbp]
    \centering
    \includegraphics[width=\textwidth]{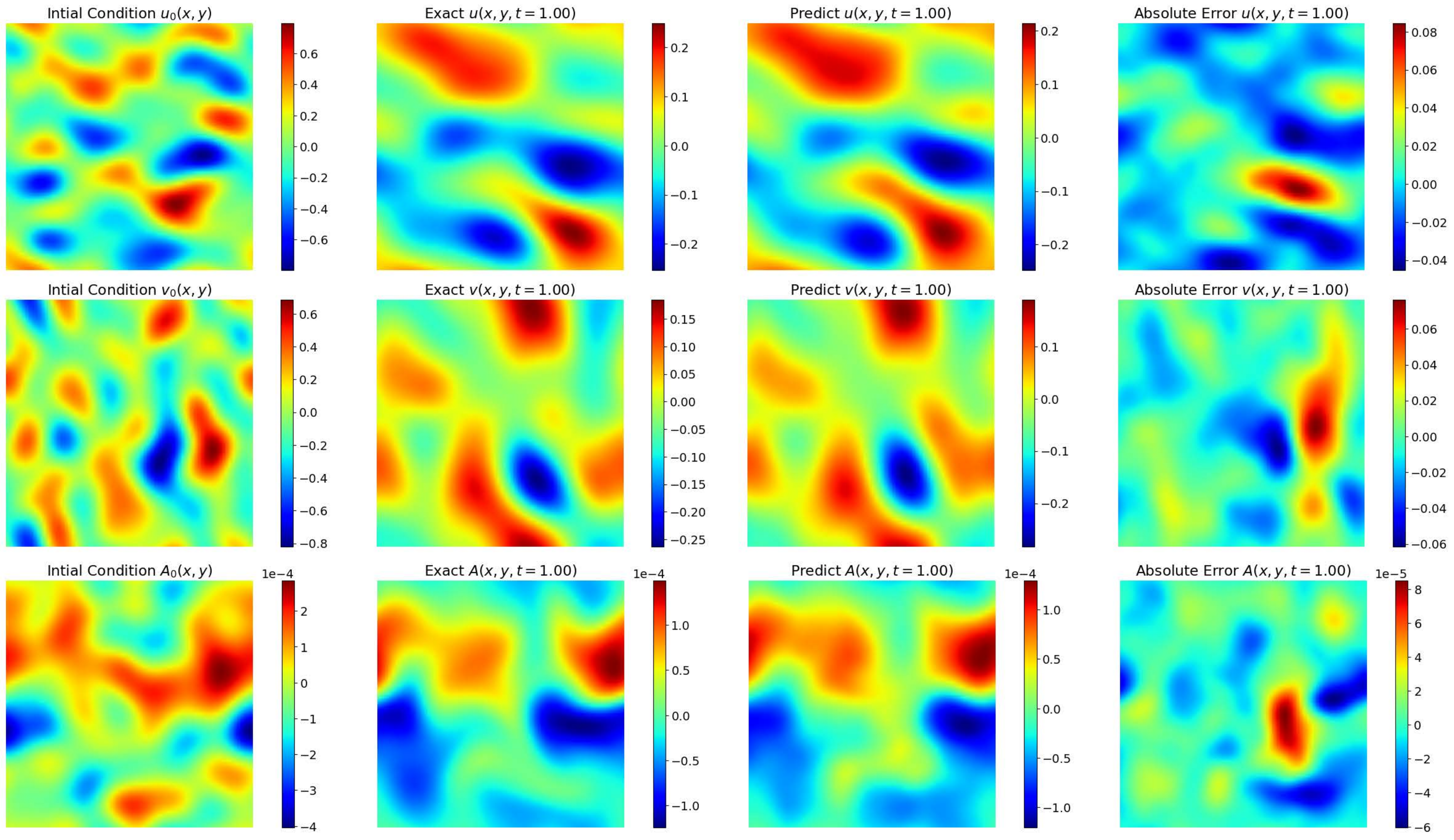}
    \caption{\textbf{$\mathbf{Re=250}$ MHD Simulations} 
    As Figure~\ref{fig:Re100_r_64} but now for a test 
    set from the $Re=250$ simulations with resolution of \(N=64^2\) grid points.}
    \label{fig:Re250_r_64}
\end{figure*}

\begin{figure*}[htbp]
    \centering
    \includegraphics[width=\textwidth]{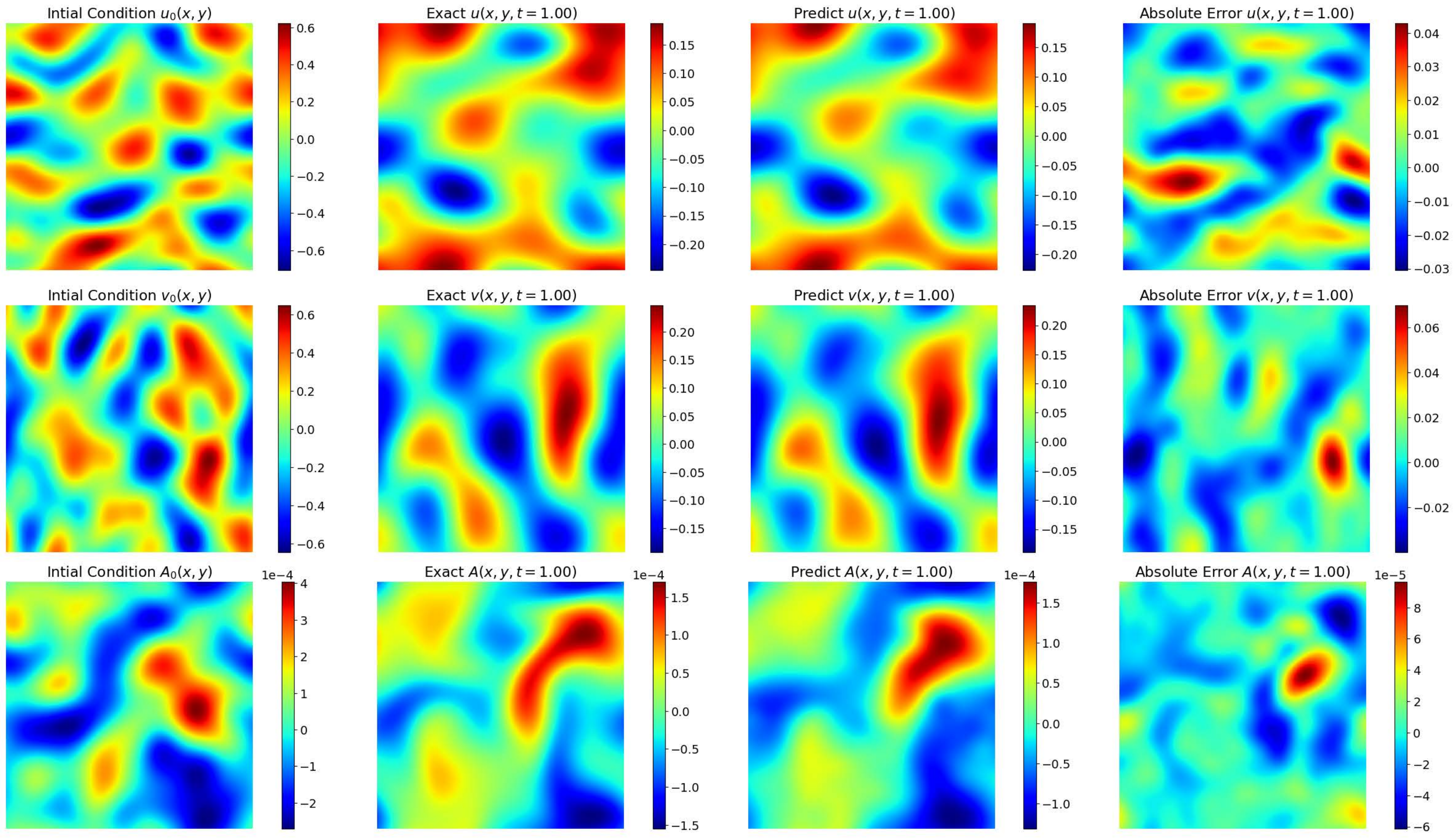}
    \caption{\textbf{$\mathbf{Re=250}$ MHD Simulations} 
    Same as Figure~\ref{fig:Re250_r_64} but now for a test 
    set with resolution of \(N=256^2\) grid points.}
    \label{fig:Re250_r_256}
\end{figure*}

\begin{figure*}[htbp]
    \centering
    \includegraphics[width=\textwidth]{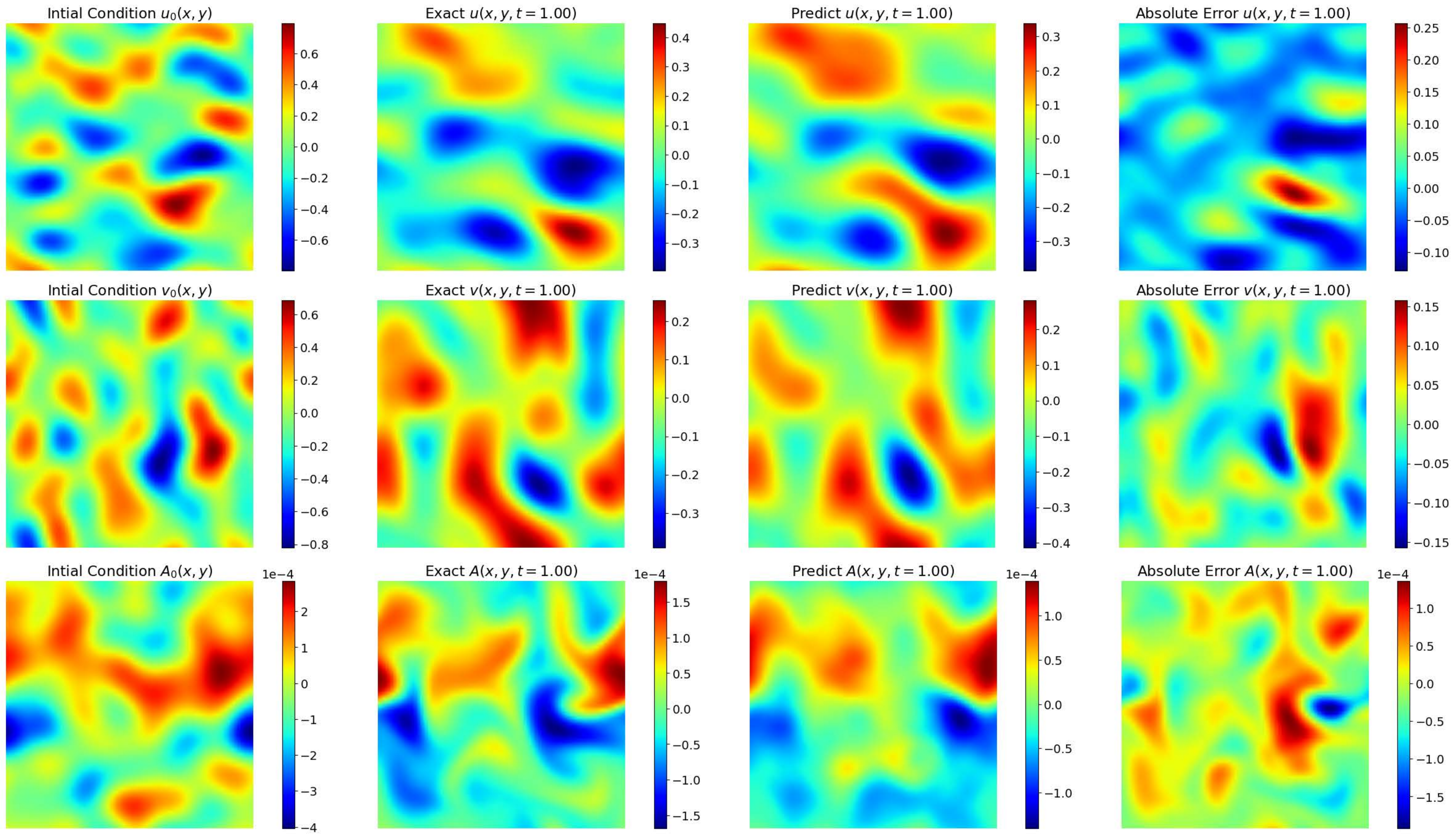}
    \caption{\textbf{$\mathbf{Re=500}$ MHD Simulations} 
    As Figure~\ref{fig:Re100_r_64} but now for a test 
    set from the $Re=500$ simulations with resolution of \(N=64^2\) grid points.}
    \label{fig:Re500_r_64}
\end{figure*}

\begin{figure*}[htbp]
    \centering
    \includegraphics[width=\textwidth]{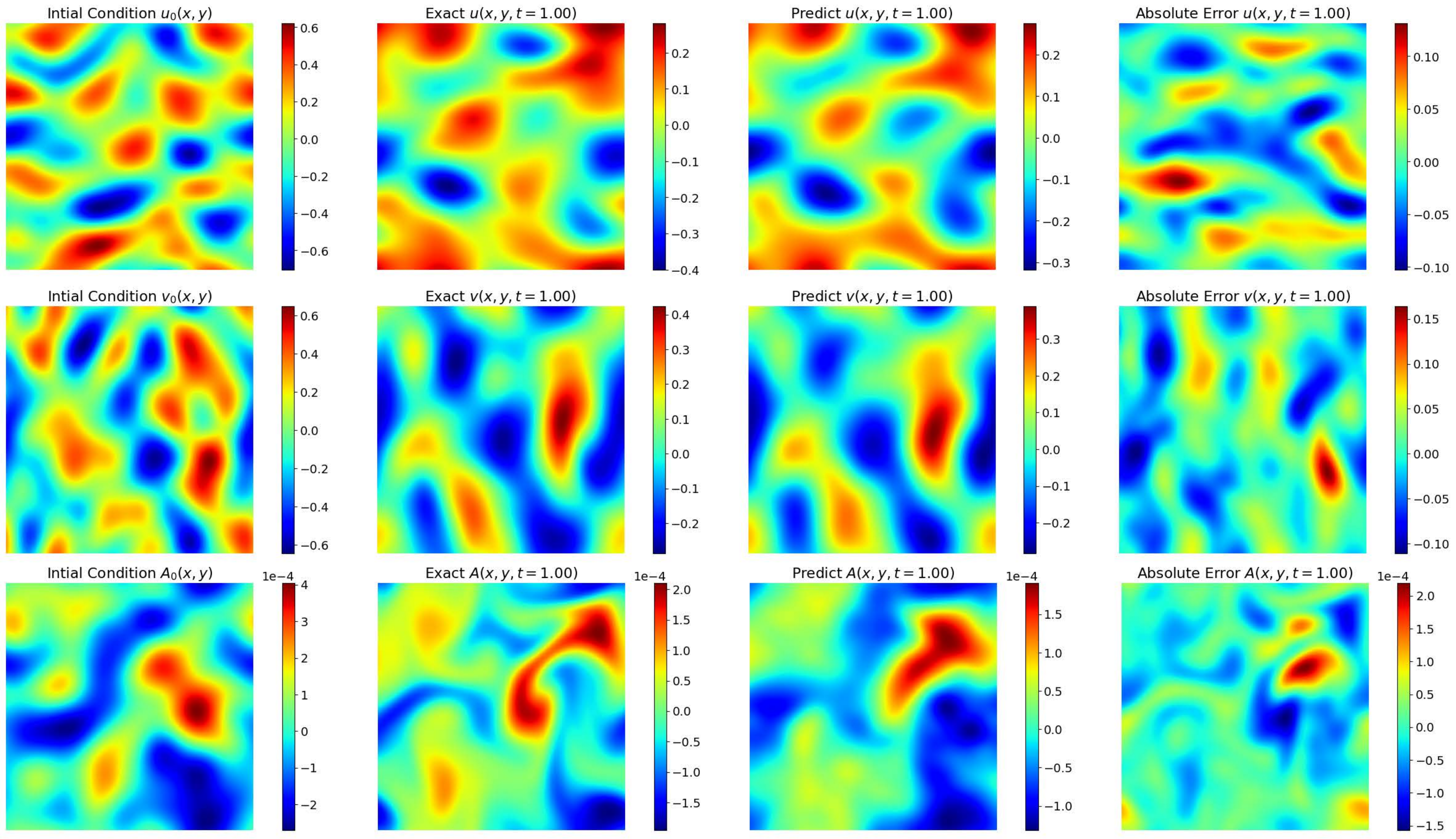}
    \caption{\textbf{$\mathbf{Re=500}$ MHD Simulations} 
    Same as Figure~\ref{fig:Re500_r_64} but now for a test 
    set with resolution of \(N=256^2\) grid points.}
    \label{fig:Re500_r_256}
\end{figure*}

\begin{figure*}[htbp]
    \centering
    \includegraphics[width=\textwidth]{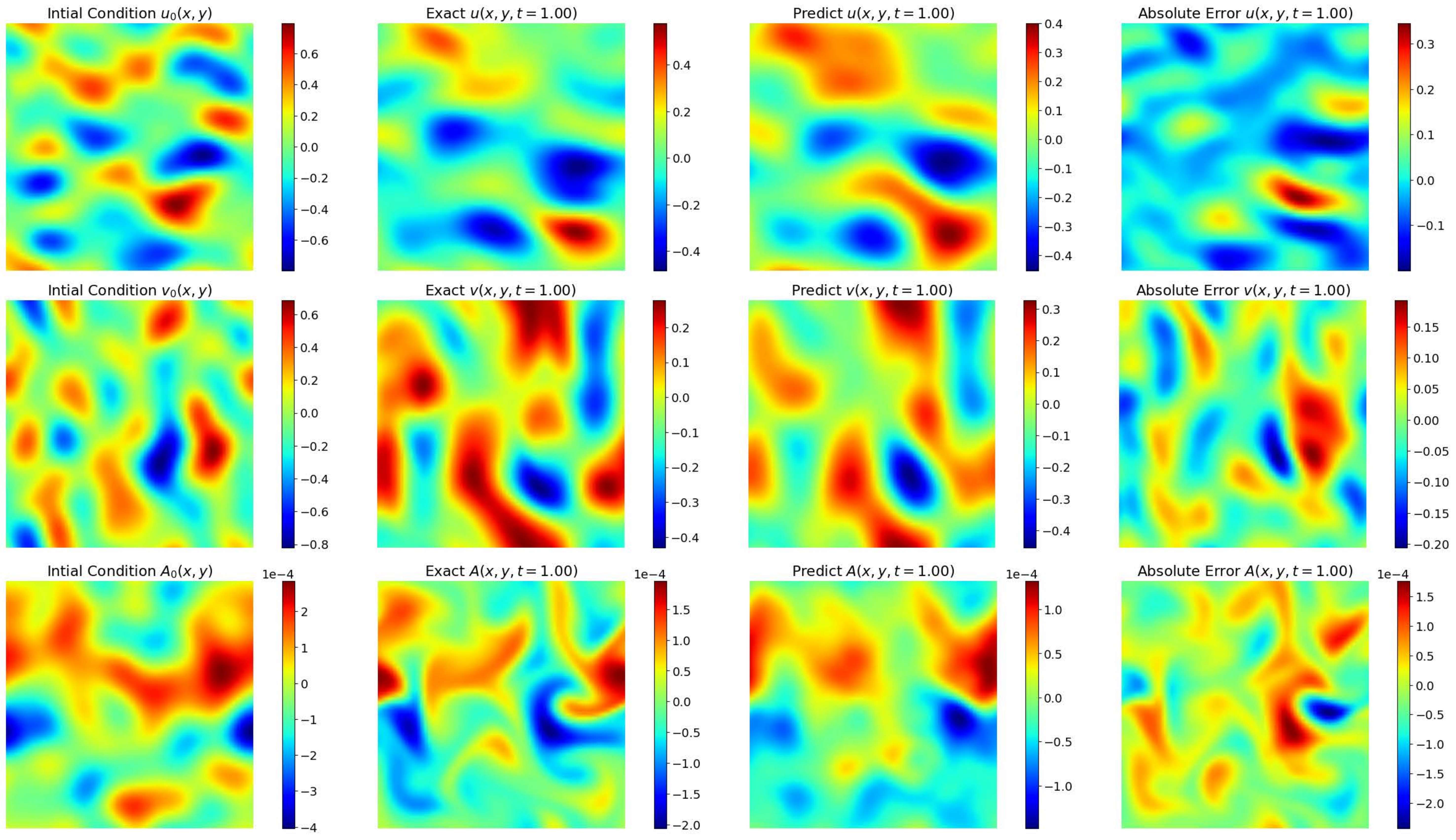}
    \caption{\textbf{$\mathbf{Re=750}$ MHD Simulations} 
    As Figure~\ref{fig:Re100_r_64} but now for a test 
    set from the $Re=750$ simulations with resolution of \(N=64^2\) grid points.}
    \label{fig:Re750_r_64}
\end{figure*}

\begin{figure*}[htbp]
    \centering
    \includegraphics[width=\textwidth]{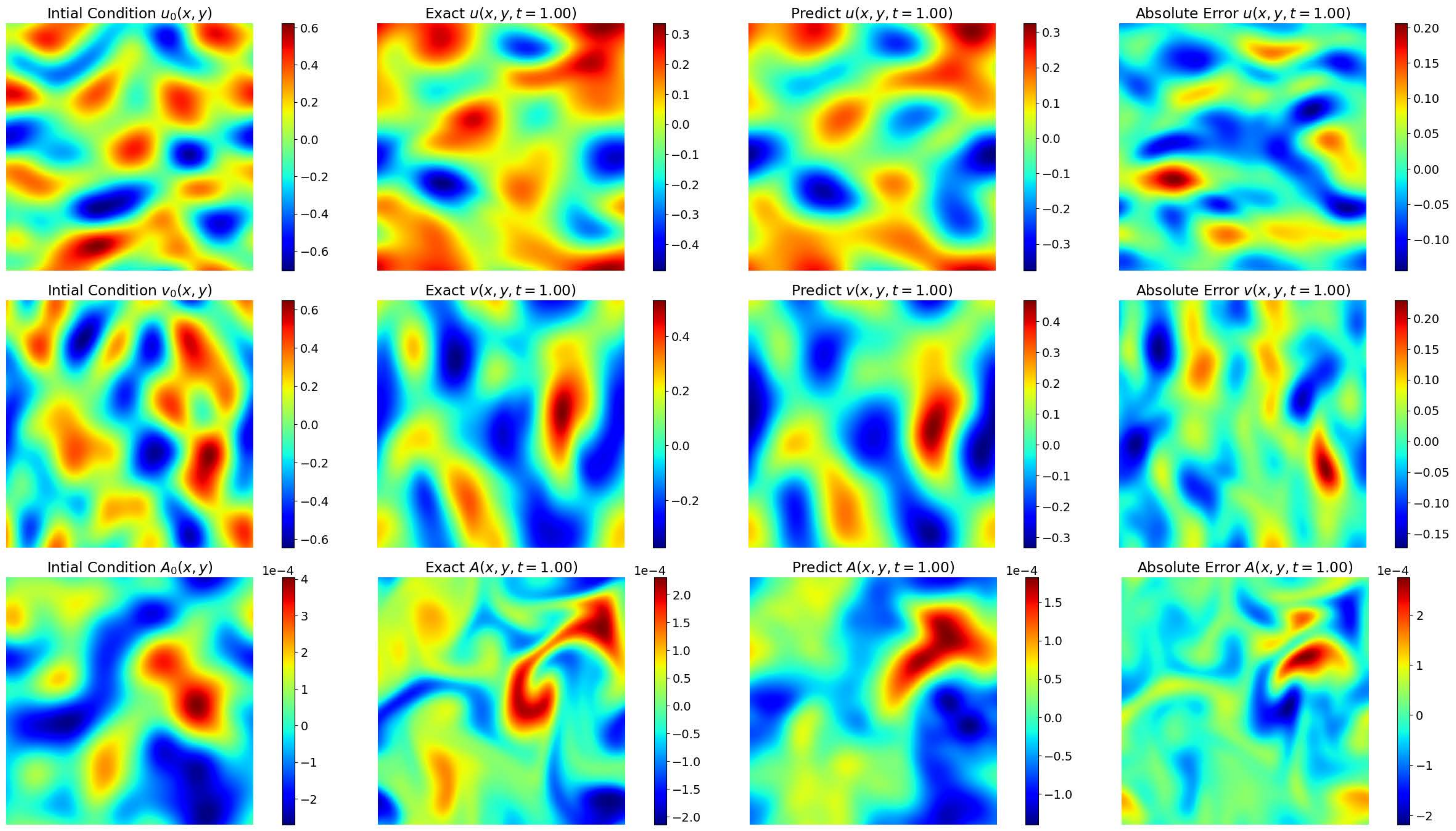}
    \caption{\textbf{$\mathbf{Re=750}$ MHD Simulations} 
    Same as Figure~\ref{fig:Re750_r_64} but now for a test 
    set with resolution of \(N=256^2\) grid points.}
    \label{fig:Re750_r_256}
\end{figure*}

\begin{figure*}[htbp]
    \centering
    \includegraphics[width=\textwidth]{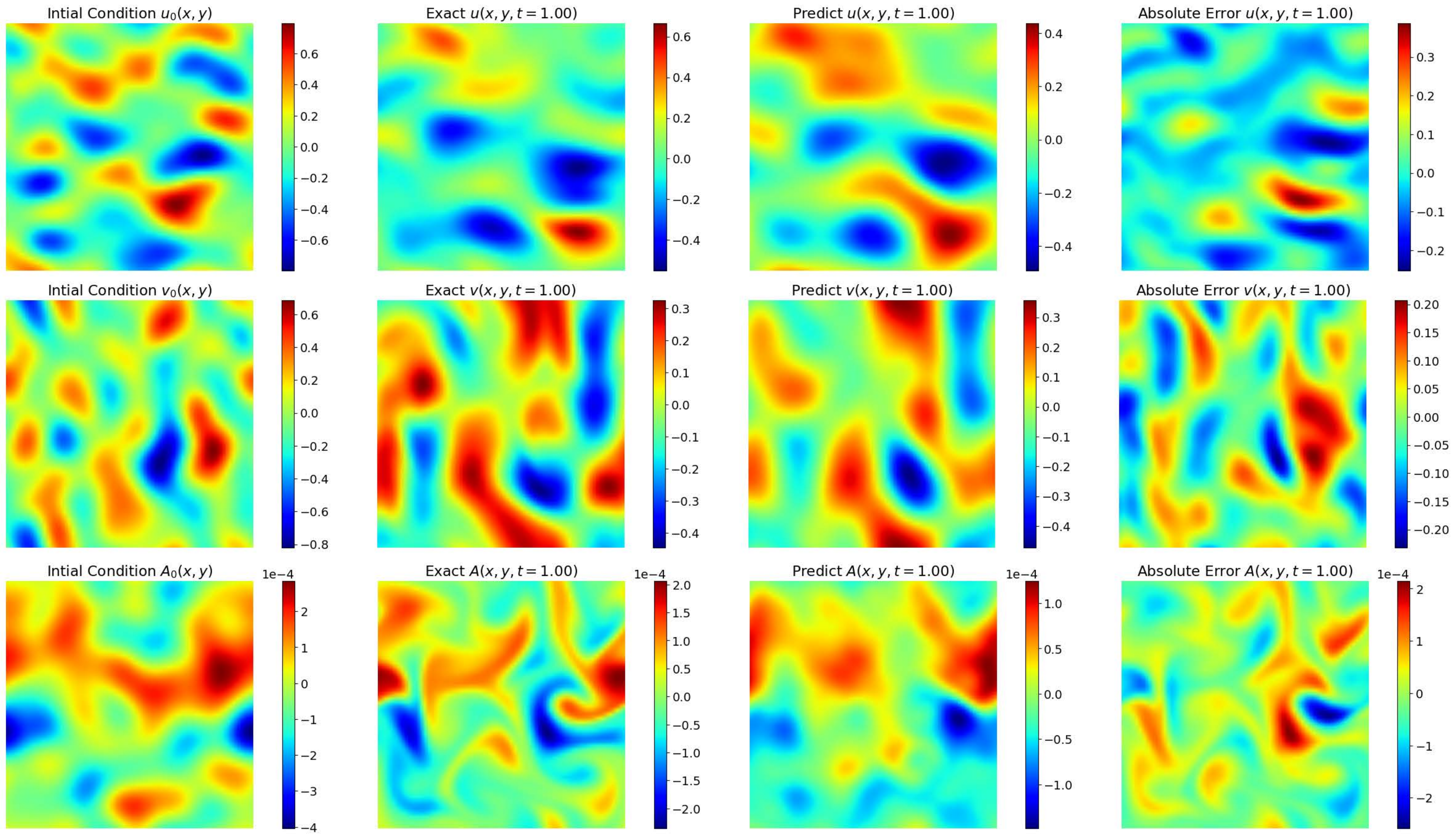}
    \caption{\textbf{$\mathbf{Re=1,000}$ MHD Simulations} 
    As Figure~\ref{fig:Re100_r_64} but now for a test 
    set from the $Re=1,000$ simulations with resolution of \(N=64^2\) grid points.}
    \label{fig:Re1000_r_64}
\end{figure*}

\begin{figure*}[htbp]
    \centering
    \includegraphics[width=\textwidth]{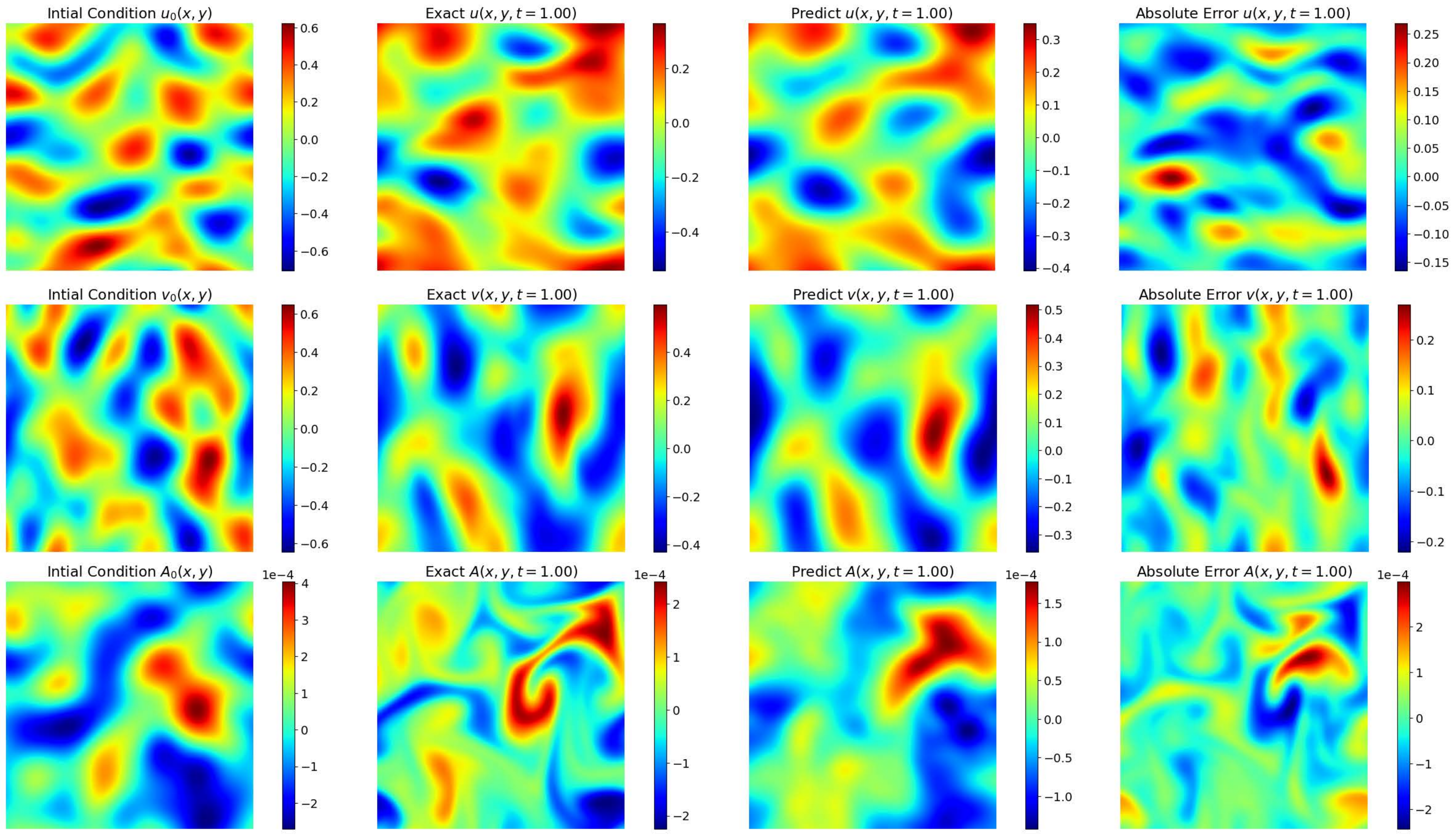}
    \caption{\textbf{$\mathbf{Re=1,000}$ MHD Simulations} 
    Same as Figure~\ref{fig:Re1000_r_64} but now for a test 
    set with resolution of \(N=256^2\) grid points.}
    \label{fig:Re1000_r_256}
\end{figure*}

Here we provide additional results that further establish 
the results we discussed in the main body of the article. 
In summary, Figures~\ref{fig:Re100_r_64}-\ref{fig:Re1000_r_256} 
show that using low, standard or high resolution simulations 
have a marginal impact in the performance of 
PINOs to capture the dynamics of turbulent MHD systems. 
What we learn from these studies is that what really matters is 
that PINOs are able to dynamically assign the correct 
kinetic and magnetic energy at high wavenumbers as the 
system evolves in time. The current approach in which 
physics-inspired loss functions remain static should be 
replaced by dynamic loss functions. This approach will 
be explored in the future.

\clearpage

\section{Additional results for low and high resolution spectra}
\label{sec:appex_spec}

\begin{figure*}[htb]
    \centering
    \includegraphics[width=\textwidth]{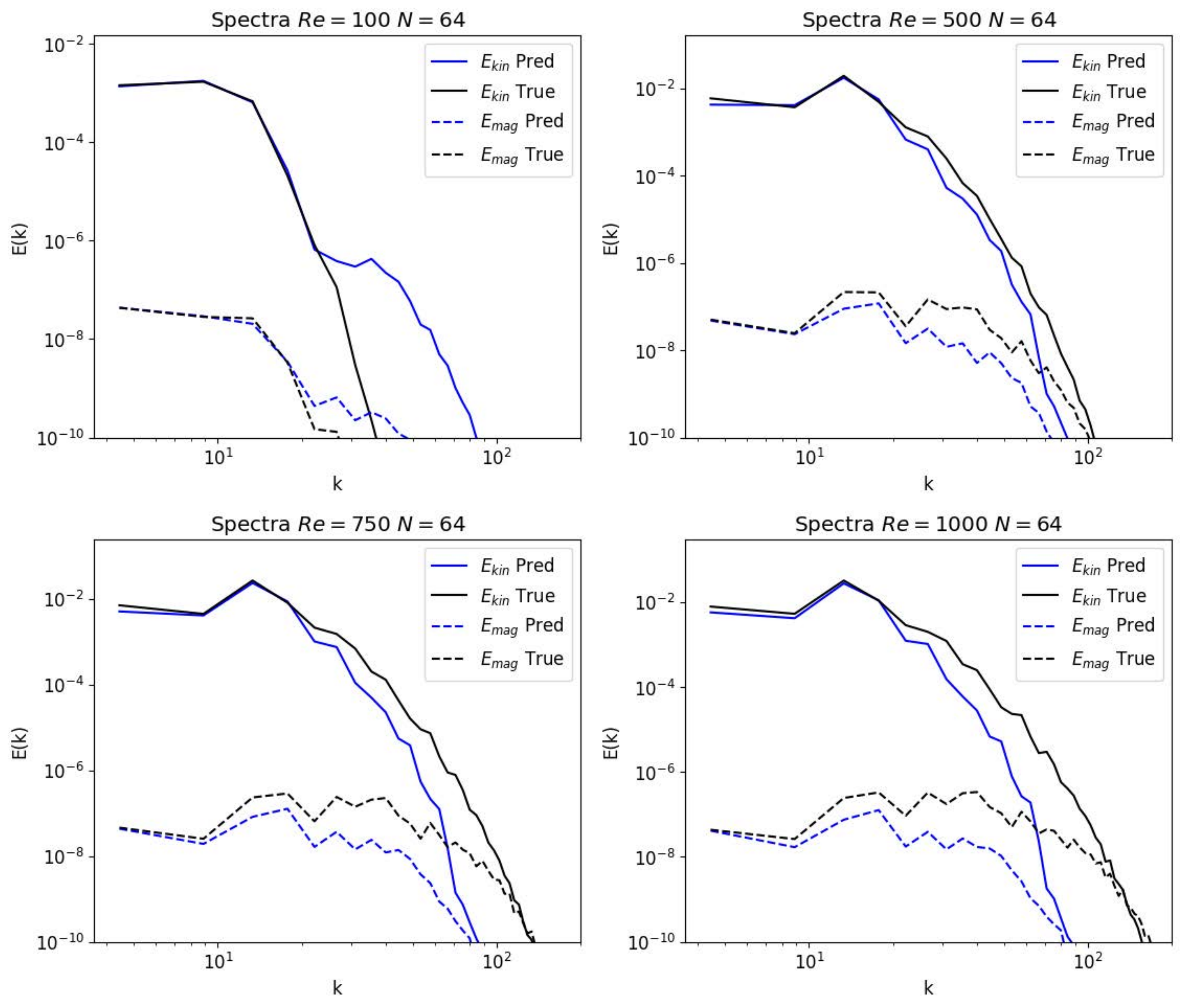}
    \caption{\textbf{Spectra at Different Resolutions} Same as Figure~\ref{ch6:fig:spectra} for the remaining $N=64^2$ models with $Re$ not plotted elsewhere.}
    \label{ch6:fig:spectra_N64}
\end{figure*}

\begin{figure*}[htb]
    \centering
    \includegraphics[width=\textwidth]{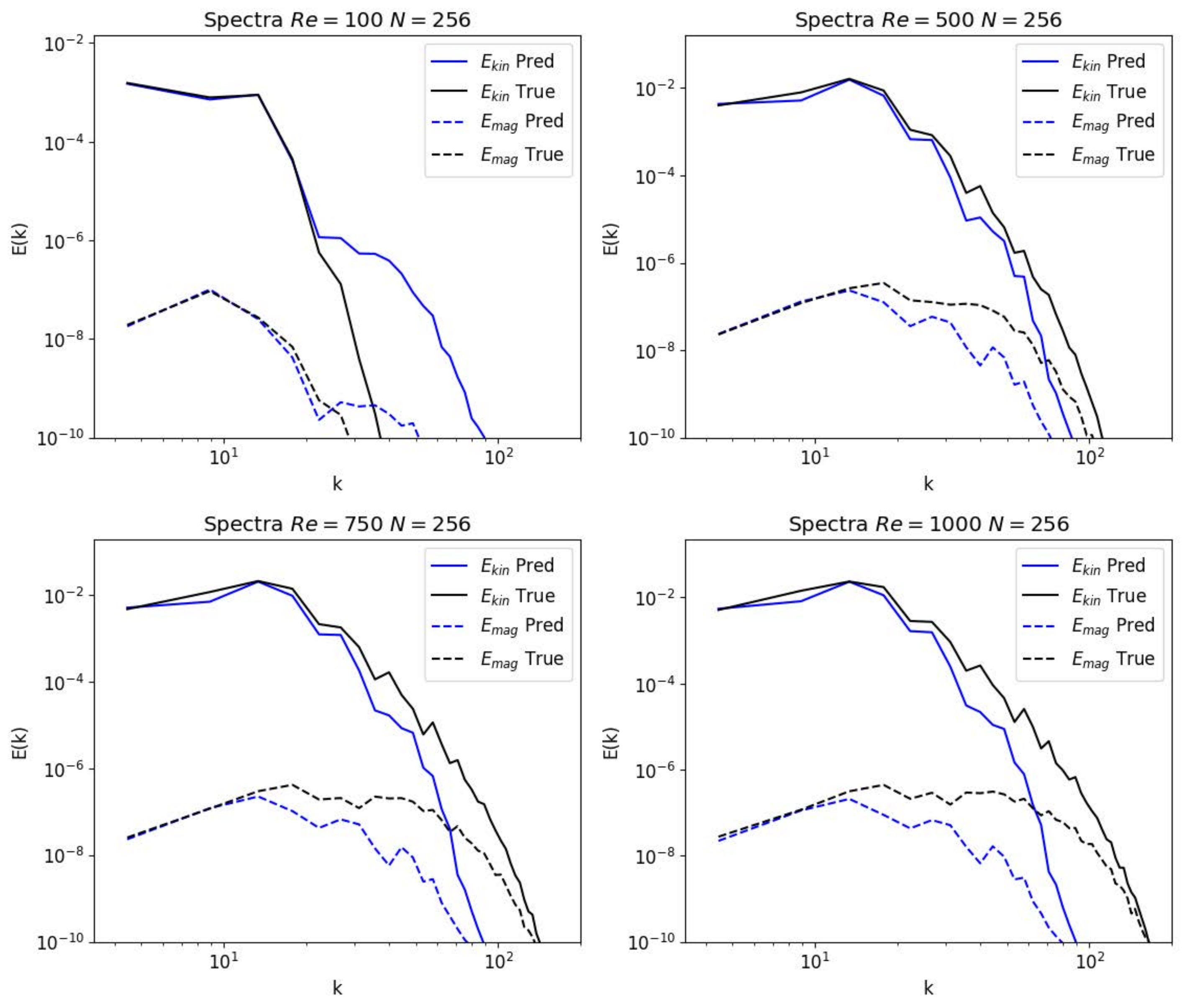}
    \caption{\textbf{Spectra at Different Resolutions} Same as Figure~\ref{ch6:fig:spectra} for the remaining $N=256^2$ models with $Re$ not plotted elsewhere.}
    \label{ch6:fig:spectra_N256}
\end{figure*}

Figures~\ref{ch6:fig:spectra_N64} and~\ref{ch6:fig:spectra_N256} 
show that low and high resolution MHD simulations 
have similar properties in their distribution of kinetic 
and magnetic energy at different wavenumbers. Thus, 
using high resolution data to train PINOs does not 
constitute a silver bullet to 
ensure that PINOs capture the small scale structure 
of turbulent flows, in particular that related to 
the impact of magnetic fields in the evolution of these 
systems.

\end{document}